\documentclass[preprint]{jpsj3}

\usepackage{bm}
\usepackage{color}
\usepackage{afterpage}
\usepackage{multicol}

\def\runauthor{Online-Journal Subcommittee}

\title{%
An Efficient Monte-Carlo Method to Make a Geometric Graph with a Fixed Connectivity\\
}

\author{
Munetaka \textsc{Sasaki}\thanks{E-mail : msasaki@kanagawa-u.ac.jp}}

\inst{
Fuculty of Engineering, Kanagawa University, Yokohama 221-8686, Japan
}

\recdate{\today}

\abst{
We present a Markov chain Monte-Carlo (MCMC) method to make a geometric graph which satisfies the following two 
conditions: (i) The degree of each vertex is fixed to a positive integer $k$. (ii) The probability that two vertices located 
on a $d$-dimensional hypercubic lattice are connected by an edge is proportional to $d_{ij}^{-\alpha}$, 
where $d_{ij}$ is the distance between the two vertices and $\alpha$ is a positive exponent. 
We introduce a reverse update method and a list-based update method for the MCMC method. 
The graph is updated efficiently by the MCMC method since the two update methods work complementarily. 
We also investigate a ferromagnetic Ising model defined on the geometric graph as a test case. 
As a result, we have confirmed that the nature of ferromagnetic transition significantly depends on the exponent $\alpha$. 
}

\kword{
Markov chain Monte-Carlo, geometric graph, ferromagnetic Ising model}

\begin{document}
\maketitle

\section{Introduction}
It is well known that thermodynamical properties of systems such as the nature of phase transition strongly 
depend on their graph structures. 
In statistical physics, there are two kinds of graphs which have been investigated intensively. 
The first one is lattice graphs in finite dimension because they are suitable to describe real systems. 
The second one is mean-field-like graphs such as fully-connected graphs and random graphs
since their analytical treatment is relatively easy. However, the thermodynamical properties of 
the two graphs are rather different in most cases. For example, although it is well-established that 
a spin glass defined on a fully-connected graph, i.e., the Sherrington-Kirkpatrick 
model~\cite{SherringtonKirkpatrick75, SherringtonKirkpatrick78}, exhibits 
a full replica-symmetry-breaking (RSB)~\cite{Parisi80}, it is very controversial whether 
a short-range spin glass on a lattice, such as a three-dimensional cubic lattice, also exhibits 
a full RSB or not. 
Therefore, it is important to investigate an {\it intermediate} graph which continuously connects 
between lattice and mean-field-like graphs. 

A way to deal with such an intermediate situation is investigating an long-range interacting system 
whose interactions algebraically decrease with increasing the distance. An example is a 
ferromagnetic Ising chain model with algebraically decreasing interactions (see Sect.~\ref{sec:result2} for details). 
In this model, we can effectively change the spatial dimension by changing an exponent $\alpha$ of the power-law decay 
of ferromagnetic interactions. When $\alpha=2$, the model exhibits a marginal Kosterlitz-Thouless 
transition~\cite{AndersonYuval71, Cardy81, Aizenman88, ImbrieNewman88, LuijtenMessingfeld01, FukuiTodo09}. 
Ising chain models with algebraically decreasing interactions have also been investigated in 
spin glasses~\cite{BrayMooreYoung86, FisherHuse88, KatzgraberYoung03, KatzgraberYoung05}. 
We can also consider a geometric graph in which the probability that two vertices are connected 
with an edge algebraically decreases with increasing the distance. This kind of geometric graph has been 
investigated not only in the field of complex networks~\cite{Kleinberg00, Coppersmith02, Biskup04} 
but also that of statistical physics like spin-glasses~\cite{Leuzzi08, Katzgraber09}.  

These two methods are, however, not appropriate in some situations. An example is a constraint 
satisfaction problem such as a graph coloring problem. Firstly, long-range systems with algebraically decreasing 
interactions are not appropriate because they are essentially different from the original constraint satisfaction problem. 
Secondly, geometric graphs, in which two vertices are connected by an edge in a stochastic manner, 
is also not appropriate if the graph is prepared in a naive way because the stochastic placement of edges naturally leads to 
a degree distribution. This means that, by varying the exponent $\alpha$ of the algebraically decreasing probabilities, 
not only the geometric property of connections among vertices but also the degree distribution are changed. 
Furthermore, statistical properties of constraint satisfaction problems are considered to strongly depend on the degree distribution. 
Therefore, even if we find some significant changes by varying $\alpha$, it is difficult to judge which is more significant reason. 

In the present study, we consider a geometric graph in which the degree of each vertex is fixed to $k$, 
where $k$ is a positive integer. Because the degree distribution is fixed, some changes caused by varying $\alpha$ 
is attributed to a change in the geometric property of connections among vertices. Since fixing 
the degree of each vertex to $k$ is a constraint condition, we can not prepare the graph in a naive way, 
i.e., we can not determine independently whether two vertices are connected by an edge or not. 
Therefore, we use a Markov chain Monte-Carlo (MCMC) method to prepare graphs. 
In the MCMC method, we introduce a reverse update method and a list-based update method. 
Because these two update methods work complementarily, we can update graphs efficiently 
by using both the methods. We also investigate a ferromagnetic Ising model defined on the geometric graph as a test case. 
As a result, it is confirmed that the nature of ferromagnetic transition significantly depends on $\alpha$.

The outline of the paper is as follows: In Sect.~\ref{sec:definition_graph}, we define our geometric graph. 
In Sect.~\ref{sec:method}, we introduce a MCMC method to make the graph. 
In Sect.~\ref{sec:result1}, we show our results on MCMC simulations to make the graph. 
Complementarity of two update methods is argued in Sect.~\ref{sec:complementarity}. 
In Sect.~\ref{sec:result2}, we show our results on a ferromagnetic Ising model defined on the geometric graph. 
Section~\ref{sec:conclusions} is devoted to conclusions. Technical details of the MCMC method are 
described in Appendixes.

\section{Definition of the Graph}
\label{sec:definition_graph}
\begin{figure}[t]
\begin{center}
\includegraphics[width=6cm]{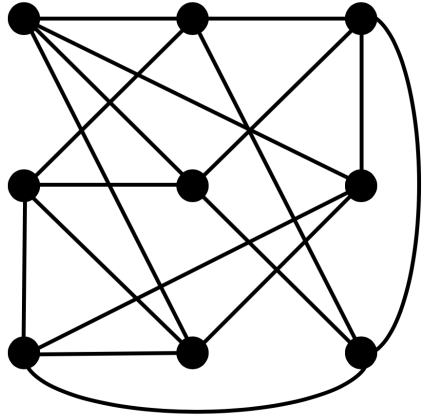}
\end{center}
\caption{An example of graph. The dimension $d$ is 2, the degree $k$ is 4, and the size $L$ is 3.}
\label{fig:graph}
\end{figure}

We consider a $d$-dimensional hypercubic lattice of size $L$. The number of lattice points $N$ is $L^d$. 
From the hypercubic lattice, we stochastically construct a graph by putting an edge between 
two lattice points $i$ and $j$ with a probability proportional to $d_{ij}^{-\alpha}$, 
where $d_{ij}$ is the distance between the two lattice points and $\alpha$ is a positive exponent. 
The distance $d_{ij}$ is measured under the periodic boundary condition. 
We also impose the following three constraint conditions:
\begin{itemize}
\item[(A)] The degree of each vertex is $k$, where $k$ is a positive integer. 
\item[(B)] There is neither multiple edge nor loop, where a loop is an edge which connects a vertex to itself. 
\item[(C)] The graph is connected, i.e., there is a path between every pair of vertices. 
\end{itemize}
Figure~\ref{fig:graph} shows an example of graph. The probability $P(G)$ for a graph $G$ to be created is given as 
\begin{equation}
P(G) = Z^{-1} \mathbb{I}(G) \prod_{(ij)\in S_{\rm E}(G) } d_{ij}^{-\alpha},
\label{eqn:Pgraph}
\end{equation}
where $Z$ is a normalization factor, $S_{\rm E}(G)$ is a set of $kN/2$ edges of the graph $G$, 
and $\mathbb{I}(G)$ is an indicator function, which one if the three constraint conditions are satisfied or zero otherwise. 
When $\alpha=0$, the graph becomes a regular random graph with a fixed connectivity 
because all the graphs which satisfy the three constraint conditions have an equal weight. 

\section{MCMC Method}
\label{sec:method}
To make graphs with a probability distribution given by Eq.~(\ref{eqn:Pgraph}), we perform the following 
two procedures:
\begin{itemize}
\item[(I)] Make an {\it initial} graph by a method which approximately satisfies 
the probability distribution of Eq.~(\ref{eqn:Pgraph}). 
\item[(II)] Change the graph by performing a MCMC simulation whose equilibrium distribution 
is given by Eq.~(\ref{eqn:Pgraph}). 
\end{itemize}
Because we perform a MCMC simulation, we can arbitrarily choose an initial graph in step (I). However, 
we make an initial graph by an approximate method and observe how well the approximate method works. 
While we update the graph by the MCMC method, we do not impose the third constraint condition (C) 
on the connectivity because the check on the connectivity is rather time-consuming. 
We check the connectivity only at the end of step (II). If the graph is connected, we finish step (II). 
If not, we perform an additional MCMC simulation until a connected graph is obtained. 
The connectivity is checked in the additional MCMC simulation at each Monte-Carlo step.

\subsection{Preparation of an initial graph}
\label{subsec:initial_graph}
\begin{figure}[t]
\begin{center}
\includegraphics[width=8cm]{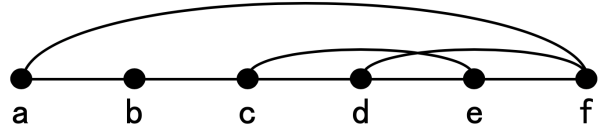}
\end{center}
\caption{An example that the procedures to prepare an initial graph with $k=3$ reach a deadlock.
To make the degree of every vertex $k$, we have to put an additional edge between $a$ and $b$. 
However, it is forbidden by the constraint condition that there is no multiple edge.  
}
\label{fig:deadlock}
\end{figure}

When we prepare an initial graph in step (I), we perform the following procedures:
\begin{itemize}
\item[(1)] Choose a vertex $i$ which does not have $k$ neighboring vertices randomly. 
\item[(2)] Set the neighboring vertices of the vertex $i$ by the following procedures:
\begin{itemize}
\item[(a)] Choose a vertex $j$ from the other $N-1$ vertices with a probability proportional to $d_{ij}^{-\alpha}$. 
\item[(b)] If the two vertices $i$ and $j$ are not connected by an edge and 
the degree of the vertex $j$ is less than $k$, connect the two vertices by an edge. 
Otherwise, return to (a).
\item[(c)] Repeat (a) and (b) until the degree of the vertex $i$ becomes $k$.
\end{itemize}
\item[(3)] Repeat (1) and (2) until all the vertices have $k$ neighboring vertices. 
\end{itemize}
In step (a), we use the Walker's alias method~\cite{Walker77}. 
In this method, we use a table of size $N-1$ to choose a vertex $j$ with ${\cal O}(1)$ computational time. 
We can use a common table for any vertex $i$ because the distance $d_{ij}$ is measured 
under the periodic boundary condition.
We therefore prepare the table at the beginning of simulation, and use it repeatedly in step (a). 
However, it should be noted that the rejection rate in step (b) increases 
with increasing the number of vertices which have $k$ neighboring vertices. 
Therefore, when the number of successive rejections in step (a) exceeds a certain number, 
we change the way to choose the neighboring vertices of vertex $i$ to a naive one, i.e., 
make a list of {\it candidates} of the neighboring vertices which do not have $k$ neighboring vertices, 
calculate the probability of each candidate which is proportional to $d_{ij}^{-\alpha}$, 
and choose the neighboring vertices from the candidates. 
We also need to pay attention to the fact that the above-mentioned procedures sometimes reach a deadlock. 
Figure~\ref{fig:deadlock} shows an example of such a case. 
When such a deadlock occurs, we again perform the preparation of an initial graph from the beginning. 

\subsection{Reverse update method}
\label{subsec:RUM}
\begin{figure}[t]
\begin{center}
\includegraphics[width=8cm]{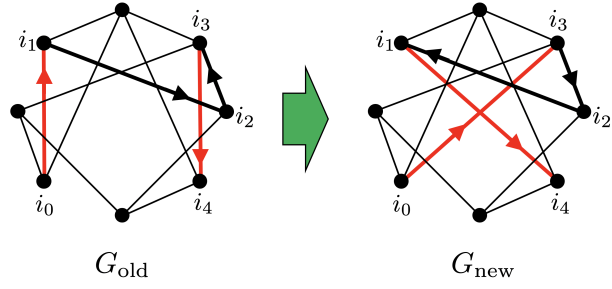}
\end{center}
\caption{(Color online) Schematic illustration of the reverse update method. The length of the path $l$ chosen in step (b) is $4$. 
After we pick up a path of length $4$ by selecting $5$ vertices $i_p$ ($p=0,1,\cdots,4$) successively, 
we change the order to visit vertices from $i_0,i_1,i_2,i_3, i_4$ to $i_0,i_3,i_2,i_1,i_4$. 
As a result, two edges $e_{i_0i_1}$ and $e_{i_3i_4}$ (red edges in $G_{\rm old}$) are removed 
and two edges $e_{i_0i_3}$ and $e_{i_1i_4}$ (red edges in $G_{\rm new}$) are added. 
}
\label{fig:RUM1}
\end{figure}

\begin{figure}[t]
\begin{center}
\includegraphics[width=8cm]{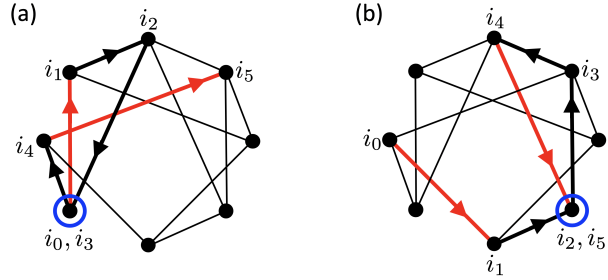}
\end{center}
\caption{(Color online) Two examples of paths which are detected in step (d). The length of the path is $5$ in both the cases. 
The path (a) has a loop which contains the starting point $i_0$ and the path (b) has a loop which contains the end point $i_5$. 
The vertices which are visited twice are enclosed by a blue circle. If we find such a loop in the path, we adopt $G_{\rm old}$ 
as $G_{\rm new}$ and stop the procedures to make a new graph $G_{\rm new}$.}
\label{fig:RUM2}
\end{figure}

We next explain the MCMC method used in step (II). Our MCMC consists of two update methods, 
i.e., reverse update method and list-based update method. 
In this subsection, we explain the reverse update method. 
This update method has also been used in the traveling salesman problem~\cite{Lin65}. 
As schematically shown in Fig.~\ref{fig:RUM1}, 
we first make a new graph $G_{\rm new}$ from the present graph $G_{\rm old}$ by the following procedures:
\begin{itemize}
\item[(a)] Choose a starting vertex $i_0$ randomly. 
\item[(b)] Choose the length of a path $l$ among $3, 4, \cdots,l_{\rm max}$ randomly with an equal probability. 
\item[(c)] Pick up a path of length $l$ by selecting $l$ vertices $i_p$ $(p=1,2,\cdots,l)$ successively.
A vertex $i_1$ is chosen among the $k$ neighboring vertices of $i_0$. 
When we choose a vertex $i_{p+1}$ $(1\le p\le l-1)$ among the neighboring vertices of $i_p$, 
we exclude $i_{p-1}$ from the candidates to prevent a back-and-forth path. 
\item[(d)] Check whether the path chosen in step (c) has a loop which contains either $i_0$ or $i_l$ 
(see Fig.~\ref{fig:RUM2}). If such a loop is detected, we adopt $G_{\rm old}$ as $G_{\rm new}$ and finish the procedures. 
Otherwise, move to the next step. 
\item[(e)] Modify the path by changing the order to visit vertices from $i_0,i_1,\cdots,i_{l-1},i_{l}$ to 
$i_0,i_{l-1},i_{l-2},\cdots, i_{1},i_{l}$. 
\end{itemize}
As we see from the procedures, the new graph $G_{\rm new}$ is made from the old one $G_{\rm old}$ 
by removing two edges $e_{i_0i_1},~e_{i_{l-1}i_{l}}$ and adding two edges $e_{i_0i_{l-1}},~e_{i_1i_{l}}$. 
Therefore, the degree of each vertex is not changed. In step (b), we impose that $l \ge 3$ to make the procedures meaningful. 
In appendix\ref{sec:appendix1}, we show that the probability 
$P_{\rm gen}^{\rm R}(G_{\rm old}\rightarrow G_{\rm new})$ that a graph $G_{\rm new}$ is generated from $G_{\rm old}$ 
satisfies the equation
\begin{equation}
P_{\rm gen}^{\rm R}(G_{\rm old}\rightarrow G_{\rm new})=P_{\rm gen}^{\rm R}(G_{\rm new}\rightarrow G_{\rm old}).
\label{eqn:Pgen_reverse}
\end{equation}
We also explain in the appendix that why the step (d) is necessary. 

The update of the graph from $G_{\rm old}$ to $G_{\rm new}$ is accepted with the probability 
of the Metropolis-Hastings method
\begin{equation}
A^{\rm R}(G_{\rm old}\rightarrow G_{\rm new})=\min\left[1, \exp(-\alpha\Delta E_{\rm graph})\right],
\label{eqn:Aprob_graph}
\end{equation}
where $\Delta E_{\rm graph}\equiv E_{\rm graph}(G_{\rm new})-E_{\rm graph}(G_{\rm old})$. 
$E_{\rm graph}(G)$ is an {\it energy} of the graph $G$ defined by
\begin{equation}
E_{\rm graph}(G)=-\log[\mathbb{I}(G)]+\sum_{(ij)\in S_{\rm E}(G)}\log(d_{ij}),
\label{eqn:def_Genergy}
\end{equation}
where $\mathbb{I}(G)$ and $S_{\rm E}(G)$ are the indicator function and 
the set of edges which appear in Eq.~(\ref{eqn:Pgraph}), respectively. 
$E_{\rm graph}(G)$ becomes $+\infty$ when $\mathbb{I}(G)$ is zero. 
We see from Eq.~(\ref{eqn:Aprob_graph}) that the exponent $\alpha$ plays a role of the inverse temperature. 
In the reverse update method, the integer $l$ chosen in step (b) determines how far the removed two edges are. 
Therefore, the energy difference $\Delta E_{\rm graph}$ tends to increase with increasing $l$. 
It is clear from Eqs.~(\ref{eqn:Pgraph}),~(\ref{eqn:Pgen_reverse}), (\ref{eqn:Aprob_graph}), and (\ref{eqn:def_Genergy})
that the transition probability 
\begin{equation}
W^{\rm R}(G_{\rm old}\rightarrow G_{\rm new})=P_{\rm gen}^{\rm R}(G_{\rm old}\rightarrow G_{\rm new})
A^{\rm R}(G_{\rm old}\rightarrow G_{\rm new}),
\end{equation}
satisfies the detailed balance condition
\begin{equation}
P(G_{\rm old})W^{\rm R}(G_{\rm old}\rightarrow G_{\rm new})=P(G_{\rm new})W^{\rm R}(G_{\rm new}\rightarrow G_{\rm old}).
\end{equation}

\begin{figure}[t]
\begin{center}
\includegraphics[width=8cm]{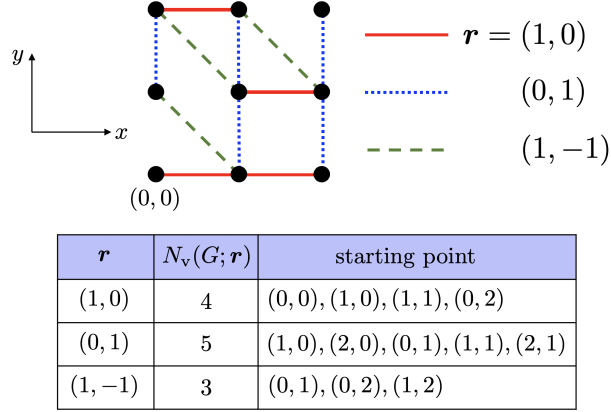}
\end{center}
\caption{(Color online) Classification of edges of a graph $G$ by their vector $\bm{r}$. In the case of 
the graph shown in the figure, edges are 
classified into the three sets which are characterized by the three vectors $(1,0)$, $(0,1)$, and $(1,-1)$.
$N_{\rm v}(G;\bm{r})$ is the number of the members of each set. The edges in each set are labeled 
by their starting points.}
\label{fig:LBUM1}
\end{figure}

\subsection{List-based update method}
\label{subsec:LB}
As shown in Fig.~\ref{fig:LBUM1}, in the list-based update method, we classify edges of a graph $G$ 
by a vector $\bm{r}_{ij}$, where $i$ and $j$ are the vertices connected 
by the edge and $\bm{r}_{ij}$ is the vector spanning from $i$ to $j$. 
We hereafter denote a set of edges of a graph $G$ for which 
either $\bm{r}_{ij}$ or $\bm{r}_{ji}$ is $\bm{r}$ by $E_{\rm v}(G; \bm{r})$. 
The number of the members of $E_{\rm v}(G;\bm{r})$ is denoted by $N_{\rm v}(G;\bm{r})$. 
The edges in $E_{\rm v}(G;\bm{r})$ are labeled by their starting points (see Fig.~\ref{fig:LBUM1}). 
We also introduce a set of vectors $V_{\rm nz}(G)$ for which $E_{\rm v}(G; \bm{r})\ne \emptyset$, 
where $\emptyset$ is the empty set. The number of the members of $V_{\rm nz}(G)$ 
is denoted by $N_{\rm nz}(G)$. In the case of Fig.~\ref{fig:LBUM1}, 
$V_{\rm nz}(G)=\{(1,0),(0,1),(1,-1)\}$ and $N_{\rm nz}(G)=3$. 
At the beginning of simulation, we make these two sets from an initial graph. 

\begin{figure}[t]
\begin{center}
\includegraphics[width=8cm]{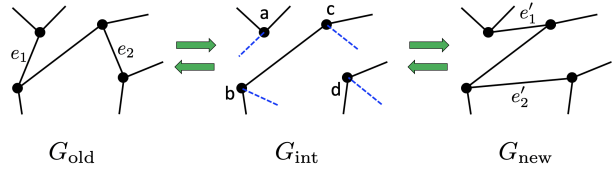}
\end{center}
\caption{(Color online) An intermediate graph $G_{\rm int}$ made by cutting the two edges $e_1$ and $e_2$ in $G_{\rm old}$ 
or $e_1'$ and $e_2'$ in $G_{\rm new}$. Blue dashed lines in $G_{\rm int}$ denote dangling edges. A graph made by 
reconnecting two vertices $a$ and $d$ firstly and $b$ and $c$ secondly is not created 
from $G_{\rm int}$ because it is forbidden by the constraint condition that there is no multiple edge. 
}
\label{fig:LBUM2}
\end{figure}

In the list-based update method, we make a new graph $G_{\rm new}$ 
from the present one $G_{\rm old}$ by the following procedures:
\begin{itemize}
\item[(a)] Choose a vector $\bm{r}_1$ from the set $V_{\rm nz}(G_{\rm old})$ with an equal probability. 
\item[(b)] Choose an edge $e_1$ from the set $E_{\rm v}(G_{\rm old}; \bm{r}_1)$. 
\item[(c)] Choose a vector $\bm{r}_2$ from the set $V_{\rm nz}(G_{\rm old}-e_1)$ with an equal probability, 
where $G-e$ is a graph made by removing the edge $e$ from the graph $G$. 
\item[(d)] Choose an edge $e_2$ from the set $E_{\rm v}(G_{\rm old}-e_1; \bm{r}_2)$.
\item[(e)] Make an intermediate graph $G_{\rm int}$ by cutting the two edges $e_1$ and $e_2$ in $G_{\rm old}$. 
As shown in Fig.~\ref{fig:LBUM2}, the graph $G_{\rm int}$ has four dangling edges. 
\item[(f)] Make a new graph $G_{\rm new}$ by reconnecting the four dangling edges. 
\end{itemize}
In step (f), we avoid making the present graph $G_{\rm old}$ if it is possible. By reconnecting the four dangling edges, 
we can make two new graphs except $G_{\rm old}$. Let us denote them by $G'$ and $G''$. 
If the both of $G'$ and $G''$ satisfy the constraint conditions mentioned in Sect.~\ref{sec:definition_graph}, we choose either $G'$ or $G''$ 
with the equal probability. If either $G'$ or $G''$ satisfies the constraint conditions, 
we choose the one which satisfies the conditions. If neither $G'$ nor $G''$ satisfies the constraint conditions, 
we choose $G_{\rm old}$ as $G_{\rm new}$. Figure~\ref{fig:LBUM2} shows the case that either $G'$ or $G''$ 
satisfies the constraint conditions. 

In the above-mentioned procedures (a)-(f), the probability $P_{\rm gen}^{\rm LB}(G_{\rm old}\rightarrow G_{\rm new})$ that 
a graph $G_{\rm new}$ is generated from $G_{\rm old}$ via an intermediate graph $G_{\rm int}$ is 
written as 
\begin{equation}
P_{\rm gen}^{\rm LB}(G_{\rm old}\rightarrow G_{\rm new})
=T_1(G_{\rm old}\rightarrow G_{\rm int})\times T_2(G_{\rm int}\rightarrow G_{\rm new}),
\label{eqn:LB1}
\end{equation}
where $T_1(G_{\rm old}\rightarrow G_{\rm int})$ is the probability that $G_{\rm int}$ is generated 
from $G_{\rm old}$ in steps (a)-(e) and $T_2(G_{\rm int}\rightarrow G_{\rm new})$ is the probability 
that $G_{\rm new}$ is generated from $G_{\rm int}$ in step (f). The probability of the reverse process 
$P_{\rm gen}(G_{\rm new}\rightarrow G_{\rm old})$ is written in a similar way. Because both the processes 
have a common intermediate graph $G_{\rm int}$ (see Fig.~\ref{fig:LBUM2}), we obtain
\begin{equation}
T_2(G_{\rm int}\rightarrow G_{\rm new})=T_2(G_{\rm int}\rightarrow G_{\rm old}).
\label{eqn:LB2}
\end{equation}
The probability $T_1(G_{\rm old}\rightarrow G_{\rm int})$ is given as 
\begin{align}
T_1(G_{\rm old}\rightarrow G_{\rm int})
&=\frac{1}{N_{\rm nz}(G_{\rm old})}\frac{1}{N_{\rm v}(G_{\rm old};\bm{r}_1)}
\frac{1}{N_{\rm nz}(G_{\rm old}-e_1)}\frac{1}{N_{\rm v}(G_{\rm old}-e_1;\bm{r}_2)}\notag\\
&+\frac{1}{N_{\rm nz}(G_{\rm old})}\frac{1}{N_{\rm v}(G_{\rm old};\bm{r}_2)}
\frac{1}{N_{\rm nz}(G_{\rm old}-e_2)}\frac{1}{N_{\rm v}(G_{\rm old}-e_2;\bm{r}_1)}.
\label{eqn:LB3}
\end{align}
The first term in the right-hand side corresponds to the case that the edge $e_1$ is chosen 
firstly and the edge $e_2$ is chosen secondly, and the second term corresponds to the case that 
the two edges are chosen in the opposite order. Similarly, the probability 
$T_1(G_{\rm new}\rightarrow G_{\rm int})$ is given as
\begin{align}
T_1(G_{\rm new}\rightarrow G_{\rm int})
&=\frac{1}{N_{\rm nz}(G_{\rm new})}\frac{1}{N_{\rm v}(G_{\rm new};\bm{r}'_1)}
\frac{1}{N_{\rm nz}(G_{\rm new}-e_1')}\frac{1}{N_{\rm v}(G_{\rm new}-e_1';\bm{r}_2')}\notag\\
&+\frac{1}{N_{\rm nz}(G_{\rm new})}\frac{1}{N_{\rm v}(G_{\rm new};\bm{r}_2')}
\frac{1}{N_{\rm nz}(G_{\rm new}-e_2')}\frac{1}{N_{\rm v}(G_{\rm new}-e_2';\bm{r}_1')},
\label{eqn:LB4}
\end{align}
where $e_1'$ and $e_2'$ are the edges which are chosen in steps (a)-(d) to make $G_{\rm int}$ from $G_{\rm new}$ 
(see Fig.~\ref{fig:LBUM2}), and $\bm{r}_1'$ and  $\bm{r}_2'$ are the corresponding vectors of $e_1'$ and $e_2'$, respectively. 

The update of the graph from $G_{\rm old}$ to $G_{\rm new}$ is accepted with the probability
\begin{equation}
A^{\rm LB}(G_{\rm old}\rightarrow G_{\rm new}) = 
\min\left[
1, \frac{T_1(G_{\rm new}\rightarrow G_{\rm int})}{T_1(G_{\rm old}\rightarrow G_{\rm int})}\exp(-\alpha\Delta E_{\rm graph})
\right].
\label{eqn:LB5}
\end{equation}
We see from Eqs.~(\ref{eqn:LB1}), (\ref{eqn:LB2}), and (\ref{eqn:LB5}) that the transition probability
\begin{equation}
W^{\rm LB}(G_{\rm old}\rightarrow G_{\rm new})=P_{\rm gen}^{\rm LB}(G_{\rm old}\rightarrow G_{\rm new})
A^{\rm LB}(G_{\rm old}\rightarrow G_{\rm new}),
\end{equation}
satisfies the detailed balance condition
\begin{equation}
P(G_{\rm old})W^{\rm LB}(G_{\rm old}\rightarrow G_{\rm new})=P(G_{\rm new})W^{\rm LB}(G_{\rm new}\rightarrow G_{\rm old}).
\end{equation}

Lastly, we roughly estimate the acceptance probability, i.e., Eq.~(\ref{eqn:LB5}). 
To this end, we first estimate $T_1(G_{\rm old}\rightarrow G_{\rm new})$ 
and $T_1(G_{\rm new}\rightarrow G_{\rm old})$ given by Eqs.~(\ref{eqn:LB3}) and (\ref{eqn:LB4}). 
Because all the graphs which appear in Eqs.~(\ref{eqn:LB3}) and (\ref{eqn:LB4}) are almost the same 
except a few edges, we can assume that $N_{\rm nz}$'s in Eqs.~(\ref{eqn:LB3}) and (\ref{eqn:LB4}) 
are almost the same. We therefore obtain
\begin{equation}
N_{\rm nz}(G)\approx N_{\rm nz}^*,
\label{eqn:LB6}
\end{equation}
where $N_{\rm nz}^*$ is a positive integer. We next suppose that the distribution of the graph is 
close to the equilibrium one given by Eq.~(\ref{eqn:Pgraph}). In this case, we can expect that 
$N_{\rm v}$'s in Eqs.~(\ref{eqn:LB3}) and (\ref{eqn:LB4}) are approximately given as
\begin{equation}
N_{\rm v}(G;\bm{r})\approx C^*|\bm{r} |^{-\alpha}, 
\label{eqn:LB7}
\end{equation}
where $C^*$ is a constant. 
By substituting Eqs.~(\ref{eqn:LB6}) and (\ref{eqn:LB7}) into Eqs.~(\ref{eqn:LB3}) and (\ref{eqn:LB4}), we obtain
\begin{align}
&T_1(G_{\rm old}\rightarrow G_{\rm int}) \approx \frac{2(|\bm{r}_1||\bm{r}_2|)^\alpha}{(N_{\rm nz}^*C^*)^2}, 
\label{eqn:LB8}\\
&T_1(G_{\rm new}\rightarrow G_{\rm int}) \approx \frac{2(|\bm{r}_1'||\bm{r}_2'|)^\alpha}{(N_{\rm nz}^*C^*)^2}.
\label{eqn:LB9}
\end{align}
The energy difference $\Delta E_{\rm graph}\equiv E_{\rm graph}(G_{\rm new})-E_{\rm graph}(G_{\rm old})$ is calculated from Eq.~(\ref{eqn:def_Genergy}) as 
\begin{equation}
\Delta E_{\rm graph}= \log(|\bm{r}_1'|)+\log(|\bm{r}_2'|)-\log(|\bm{r}_1|)-\log(|\bm{r}_2|),
\label{eqn:LB10}
\end{equation}
where we have assumed that both $\mathbb{I}(G_{\rm old})$ and $\mathbb{I}(G_{\rm new})$ are $1$. 
By substituting Eqs.~(\ref{eqn:LB8}), (\ref{eqn:LB9}), and (\ref{eqn:LB10}) into Eq.~(\ref{eqn:LB5}), we obtain
\begin{equation}
A^{\rm LB}(G_{\rm old}\rightarrow G_{\rm new}) \approx 1.
\label{eqn:LB11}
\end{equation}
This means that the list-based update method becomes effectively rejection-free. 
However, it should be noted that this result crucially relies on the approximation of Eq.~(\ref{eqn:LB7}). 
In appendix\ref{sec:appendix2}, we show that Eqs.~(\ref{eqn:LB7}) and (\ref{eqn:LB11}) are not valid when $\alpha \ge d$.

\section{Result I: MCMC Simulation for Graph Construction}
\label{sec:result1}
In this section, we show results of MCMC simulations for graph construction. 
We set the dimension $d$ to $1$ and mainly investigate the case that the exponent $\alpha$ is $2$. 
This case is important because the ferromagnetic Ising model on the graph is expected 
to exhibit a marginal Kosterlitz-Thouless transition (see Sect.~\ref{sec:result2} for details). 
In the simulation, we first prepare an initial graph by the method mentioned in Sect.~\ref{subsec:initial_graph}. 
Then, the graph is updated by the following three different methods: 
\begin{itemize}
\item [(a)] Reverse update method. 
\item [(b)] List-based update method. 
\item [(c)] Both the reverse and list-based update methods. 
\end{itemize}
In the methods (a) and (b), one Monte-Carlo step (MCS) is defined by $N_{\rm edge}$ trials 
to choose two edges and reconnect them, where $N_{\rm edge} = Nk/2$ is the number of edges. 
In the method (c), one MCS is defined by $N_{\rm edge}$ trials of the reverse update method and subsequent 
$N_{\rm edge}$ trials of the list-based update method. 

\begin{figure}[t]
\begin{center}
\includegraphics[width=8cm]{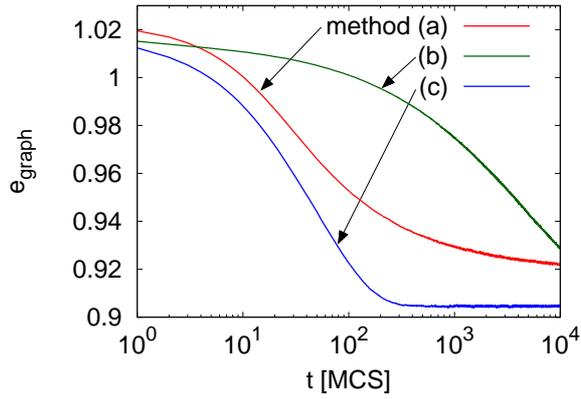}
\vspace{5mm}
\end{center}
\caption{(Color online) Time dependence of $e_{\rm graph}$, 
where $e_{\rm graph}\equiv E_{\rm graph}/N_{\rm edge}$, $E_{\rm graph}$ is defined by Eq.~(\ref{eqn:def_Genergy}), 
and $N_{\rm edge}=Nk/2$ is the number of edges. 
The lattice size $L$ is $300,000$, the dimension $d$ is $1$, the degree $k$ is $4$, 
and the exponent $\alpha$ is $2$. The maximum path length $l_{\rm max}$ 
in the reverse update method is $10$. The average is taken over $100$ different graphs.}
\label{fig:EneRelax}
\end{figure}

\begin{figure}[t]
\begin{center}
\includegraphics[width=8cm]{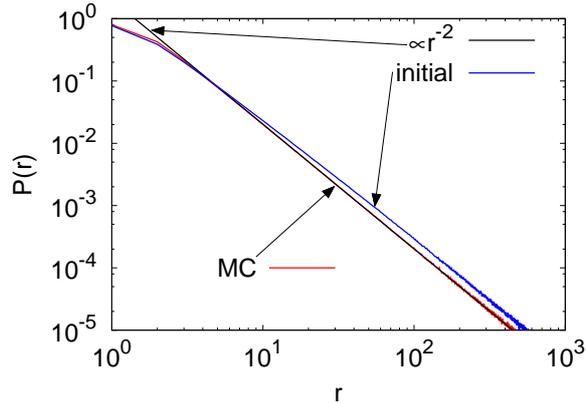}
\vspace{5mm}
\end{center}
\caption{(Color online) The probability distribution $P(r)$ that two vertices separated by a distance $r$ 
are connected by an edge is plotted as a function of $r$. $P(r)$ for initial graph is denoted by blue line and 
that after the graph is updated for $10,000$~MCS by the method (c) is denoted by red line. 
The black straight line is proportional to $r^{-2}$. 
The lattice size $L$ is $300,000$, the dimension $d$ is $1$, the degree $k$ is $4$, 
and the exponent $\alpha$ is $2$. The maximum path length $l_{\rm max}$ 
in the reverse update method is $10$. The average is taken over $100$ different graphs.}
\label{fig:Hist}
\end{figure}

Figure~\ref{fig:EneRelax} shows the time dependence of the energy defied by Eq.~(\ref{eqn:def_Genergy}).
The lattice size $L$ is $300,000$. 
We clearly see that the energy relaxes to an equilibrium value. This means that 
there is a distinct difference between the initial distribution realized by the approximate method 
described in Sect.~\ref{subsec:initial_graph} and the equilibrium distribution given by Eq.~(\ref{eqn:Pgraph}). 
We also notice that the relaxation speeds of the three methods are quite different. 
When we use the reverse update method or the list-based update method individually, 
the relaxation speed is rather slow. The energy does not relax to the equilibrium value 
within $10,000$~MCS in both the cases. However, when we use both the methods, the energy reaches to the 
equilibrium value after a few hundred MCS. This result indicates that the two update methods work complementarily. 

In Fig.~\ref{fig:Hist}, we show the probability distribution $P(r)$ that two vertices separated 
by a distance $r$ are connected by an edge. We measure $P(r)$ after 
the graph is updated for $10,000$~MCS by the method (c). We also measure 
$P(r)$ for initial graphs for comparison. We see a small but distinct difference between them. 
The relaxation of the energy shown in Fig.~\ref{fig:EneRelax} is caused by this difference. 

\begin{figure}[t]
\begin{center}
\includegraphics[width=8cm]{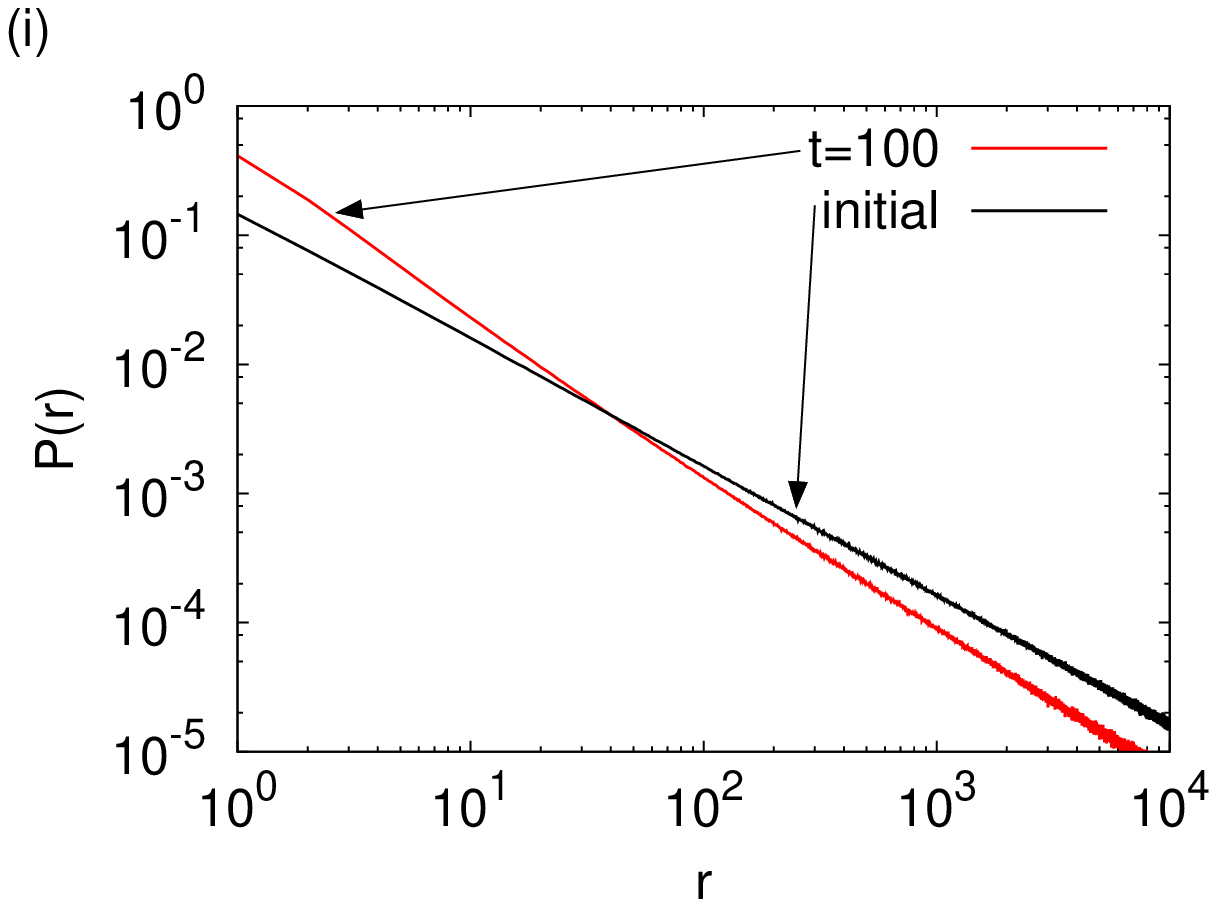}
\end{center}
\begin{center}
\includegraphics[width=8cm]{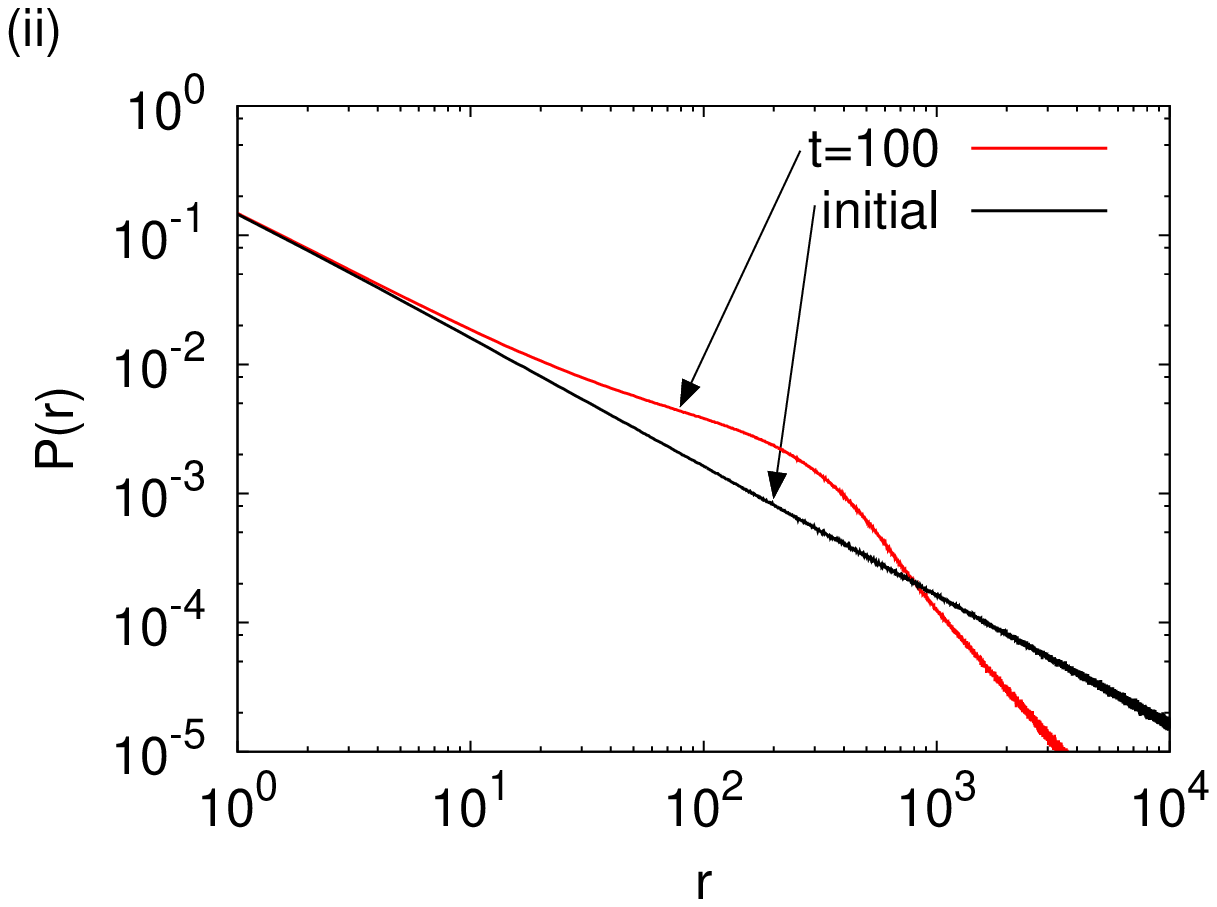}
\end{center}
\vspace{10mm}
\caption{(Color online) The probability distribution $P(r)$ after the graph is updated by 
either the method (a) (panel (i)) or (b) (panel (ii)) for $100$~MCS is 
plotted as a function of $r$ (red lines). $P(r)$ for initial graph, 
which is prepared with $\alpha=1$, is also shown for comparison (black lines). 
In the MCMC simulation, $\alpha$ is set to $2$. 
The lattice size $L$ is $300,000$, the dimension $d$ is $1$, and the degree $k$ is $4$. 
The maximum path length $l_{\rm max}$ in the reverse update method is $10$. 
The average is taken over $100$ different graphs.}
\label{fig:Pchange}
\end{figure}

To see how the two update methods work complementarily, 
we measure the probability distribution $P(r)$ after the graph is updated by 
either the method (a) or (b) for $100$~MCS. We set $\alpha$ to $2$ in the MCMC simulation. 
However, when we prepare initial graphs, we set $\alpha$ to $1$ to clearly observe 
the change in $P(r)$. Figure~\ref{fig:Pchange} shows the result. 
We also show $P(r)$ for initial graph for comparison. We see from Fig.~\ref{fig:Pchange}~(i) that 
the reverse update method mainly changes $P(r)$ for small $r$. 
On the other hand, the list-based update method mainly changes $P(r)$ for large $r$. 
As a consequence, the exponent for large $r$ changes to $2$ (see Fig.~\ref{fig:Pchange}~(ii)). \
This reason why $P(r)$ for large $r$ is changed by the list-based update method is naturally 
understood from the fact that a vector $\bm{r}$ to which an edge belongs is chosen 
with an equal probability (see the steps (a) and (c) in the list-update method).

\begin{figure}[t]
\begin{center}
\includegraphics[width=8cm]{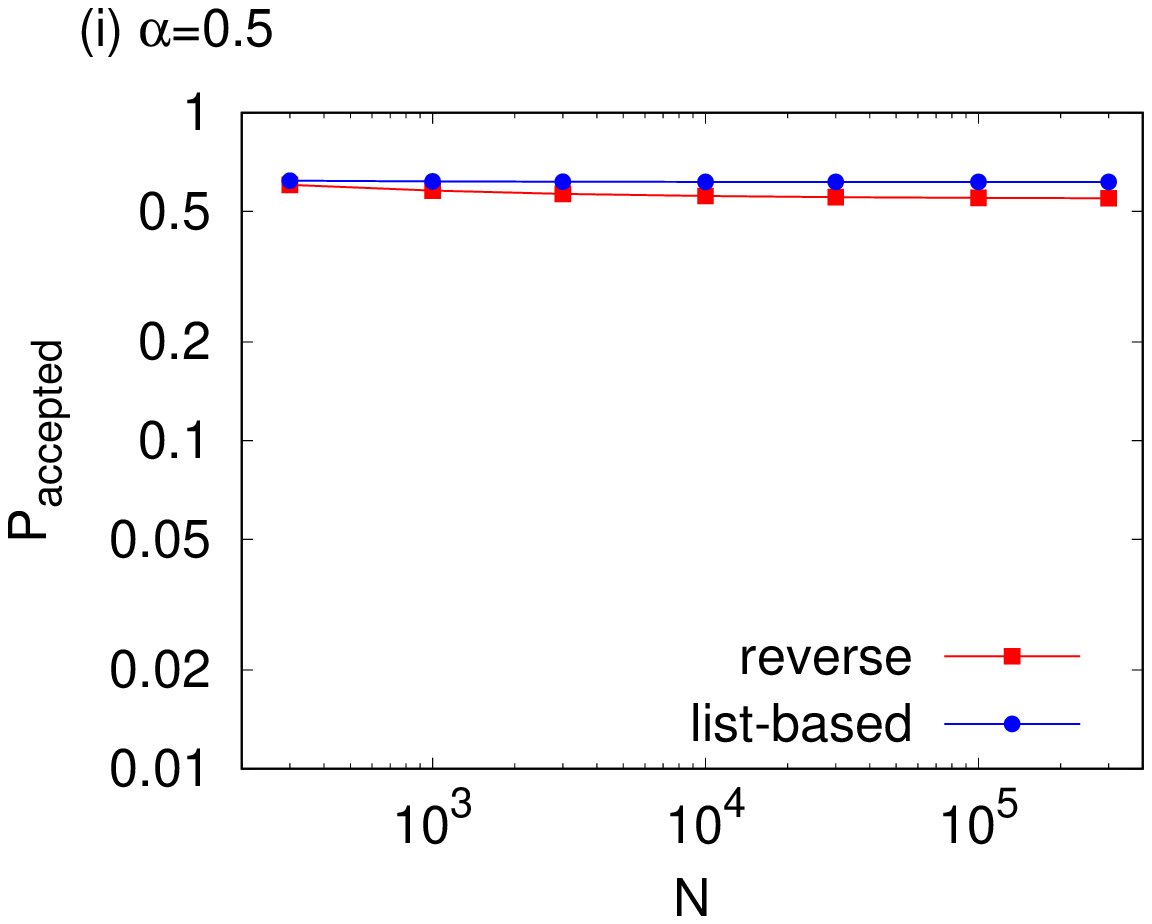}
\end{center}
\begin{center}
\includegraphics[width=8cm]{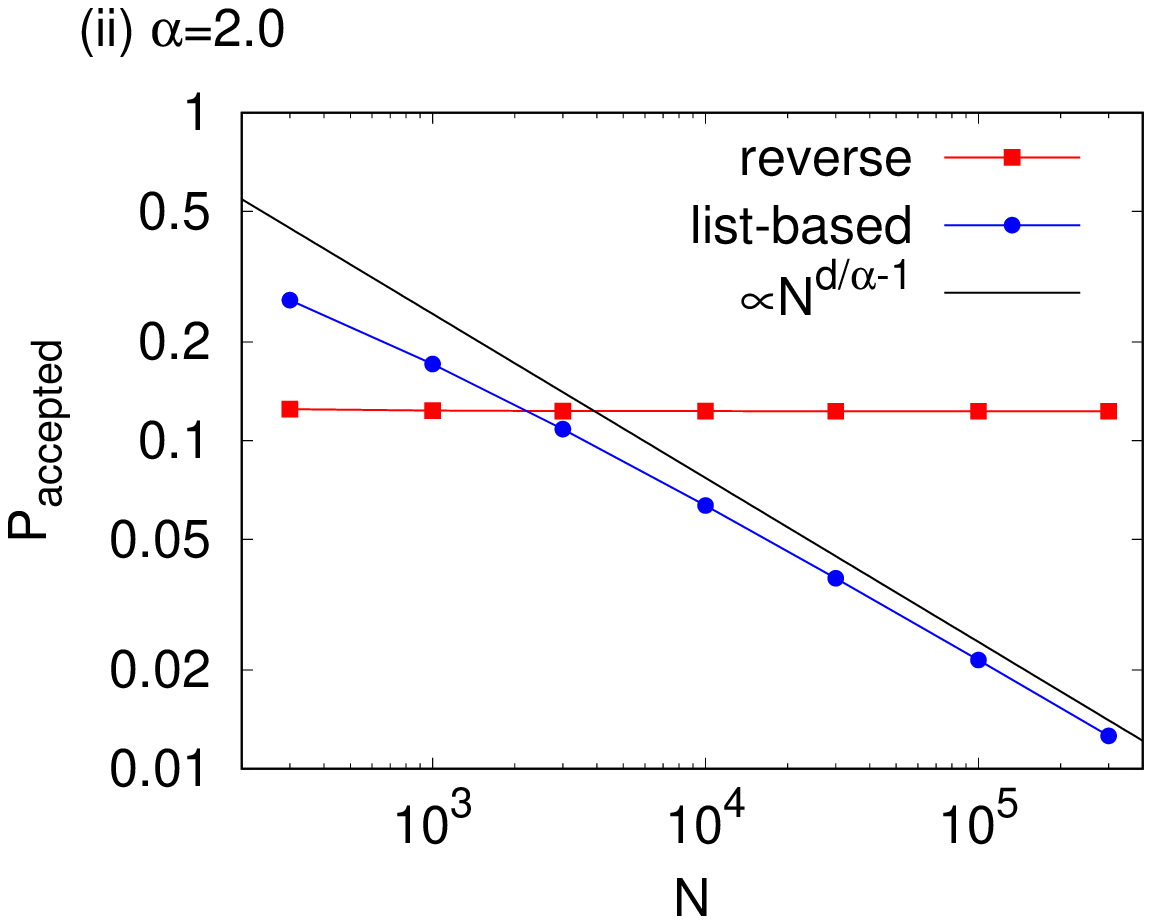}
\end{center}
\vspace{10mm}
\caption{(Color online) The acceptance probability of the two methods for (i) $\alpha=0.5$ and 
(ii) $\alpha=2.0$ is plotted as a function of the number of vertices $N$. 
The dimension $d$ is $1$ and the degree $k$ is $4$. 
The black straight line in the panel (ii) is proportional to $N^{d/\alpha-1}$.
The maximum path length $l_{\rm max}$ in the reverse update method is $10$. 
 The acceptance probability is measured after the graph is updated for $10,000$~MCS by the method (c).
The average is taken over $100$ different graphs.}
\label{fig:Paccepted}
\end{figure}

In Fig.~\ref{fig:Paccepted}, we show the size dependence of the acceptance probability of the two methods 
for (i) $\alpha=0.5$ and (ii) $\alpha=2.0$. The acceptance probability of the reverse update method 
hardly depends on the size in both the cases (i) and (ii). On the contrary,  the acceptance probability of 
the list-based update method decreases with increasing the size when $\alpha=2.0$. 
We show in appendix\ref{sec:appendix2} that the acceptance probability of the list-based update method 
for $\alpha > d$ is proportional to $N^{d/\alpha-1}$. This means that the acceptance probability 
is zero in the thermodynamic limit $N\rightarrow \infty$ when $\alpha > d$. 
The reason why the acceptance probability for $\alpha > d$ decreases with increasing the size 
is also explained in appendix\ref{sec:appendix2}.

\begin{figure}[t]
\begin{center}
\includegraphics[width=8cm]{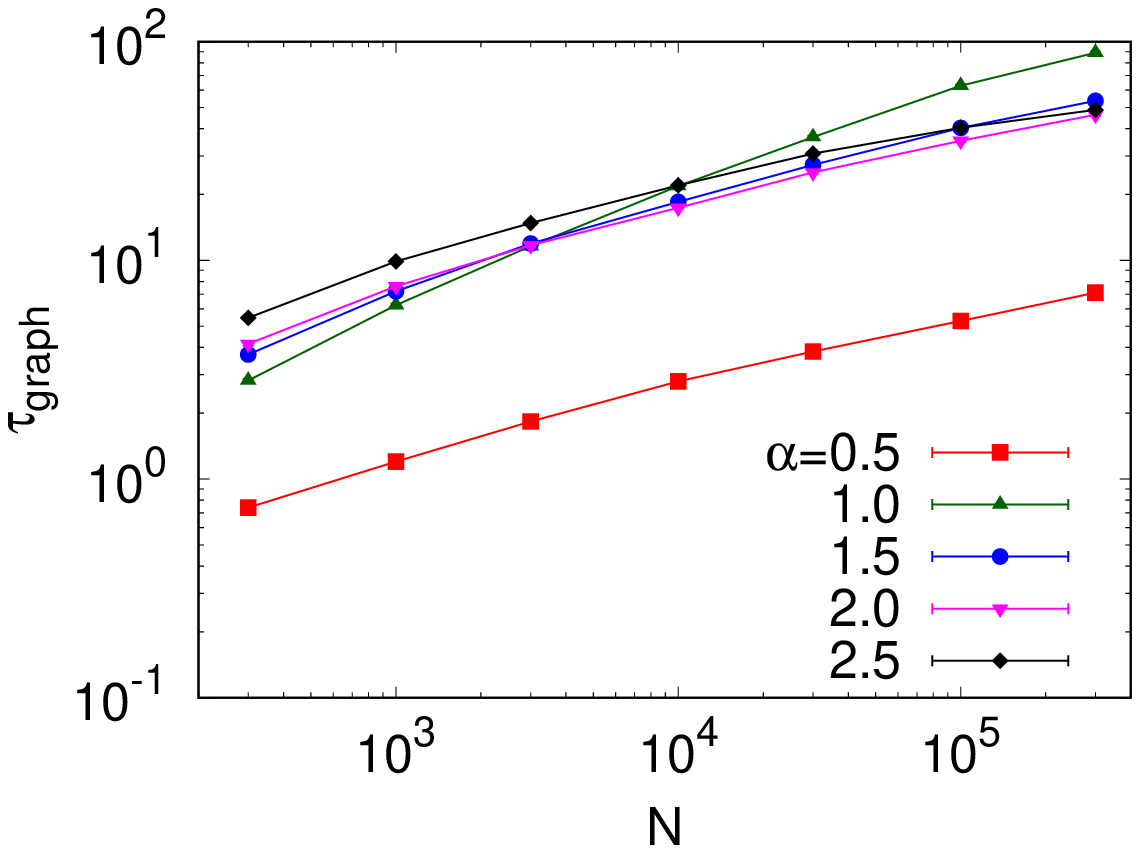}
\end{center}
\vspace{10mm}
\caption{(Color online) The autocorrelation time $\tau_{\rm graph}$ of $E_{\rm graph}$ is plotted 
as a function of the number of vertices $N$ for $\alpha=0.5$ (squares), $1.0$ (triangles), 
$1.5$ (circles), $2.0$ (upside-down triangles), and $2.5$ (diamonds). 
The dimension $d$ is $1$ and the degree $k$ is $4$. The graph is updated by the method (c). 
The maximum path length $l_{\rm max}$ in the reverse update method is $10$. 
$\tau_{\rm graph}$ is measured after the graph is updated for $10,000$~MCS. 
The average is taken over $100$ different graphs.}
\label{fig:RelaxTime}
\end{figure}

Lastly, we show the size dependence of the autocorrelation time $\tau_{\rm graph}$ of $E_{\rm graph}$ 
in Fig.~\ref{fig:RelaxTime}. The autocorrelation time is measured by a method in Ref.~\citen{autocorrelation_time}. 
The graph is updated by the method (c). We see that the increase in $\tau_{\rm graph}$ with increasing the size 
is rather gradual. The autocorrelation time is less than $100$~MCS even when $N =300,000$. 
We also find that $\tau_{\rm graph}$ weakly depends on $\alpha$ when $\alpha > d =1$. 
Because the acceptance probability of the list-based update method for $\alpha > d$ is proportional to $N^{d/\alpha-1}$, 
one may consider that $\tau_{\rm graph}$ rapidly increases with increasing $\alpha$. 
However, such a tendency is not observed from the data. 
On the contrary, when $N$ is large, $\tau_{\rm graph}$ with $\alpha=1$ is larger than that with $\alpha > 1$, 
although the acceptance probability of the list-based update method for $\alpha=1$ slowly decreases 
with increasing $N$. We have also measured how $\tau_{\rm graph}$ depends on $\alpha$ 
when the graph is updated by either the method (a) or (b), and found that $\tau_{\rm graph}$ 
steadily increases with increasing $\alpha$ in both the cases (the data are not shown). 
Therefore, it is not easy to understand why the autocorrelation time weakly depends on $\alpha$ only in the method (c). 
Anyhow, this weak dependence of $\tau_{\rm graph}$ on $\alpha$ in the method (c) is a good result 
from a practical point of view.

\section{Complementarity of the Two Update Methods}
\label{sec:complementarity}
As we see in the previous section, the two update methods have their own drawbacks. 
A problem of the reverse update method is that it slowly changes the distribution of $P(r)$ 
for large $r$ (see Fig.~\ref{fig:Pchange}~(i)). This is the reason why the energy of the method~(a) 
in Fig.~\ref{fig:EneRelax} is far from the equilibrium value even after $10,000$~MCS. 
On the other hand, an apparent drawback of the list-based update method is that the acceptance probability 
decays with increasing the size when $\alpha > d$ (see Fig.~\ref{fig:Paccepted}~(ii)). 
Therefore, the energy of the method~(b) also does not reach to the equilibrium value within $10,000$~MCS. 
However, the acceptance probability of the list-based update method is not so small even for $N=300,000$ 
because the absolute value of the exponent of the power-law decay $|d/\alpha-1|$ is not large. 
Furthermore, because the list-based update method mainly changes $P(r)$ 
for large $r$ (see Fig.~\ref{fig:Pchange}~(ii)), it makes up the drawback of the reverse update method. 
Therefore, as shown in Fig.~\ref{fig:EneRelax}, the relaxation speed is much improved by using 
both the update methods.  

We point out that the list-based update method also makes up a drawback 
of the reverse update method on the ergodicity. Because the length of the path 
chosen in the reverse update method is finite, it is not clear whether the reverse update method 
is ergodic or not. At least, the reverse update method is not ergodic if we do not 
impose the third constrained condition (C) on the connectivity of the graph. 
Because the reverse update method choose two edges which are connected by a path, 
it does not change the connectivity of the graph. Therefore, if a graph is disconnected, 
it is never updated to a connected graph by the reverse update method. 
However, the list-based update method is ergodic because there is a possibility that 
any two edges are chosen. The two update methods are again complementary 
from this point of view. 

\begin{figure}[t]
\begin{center}
\includegraphics[width=8cm]{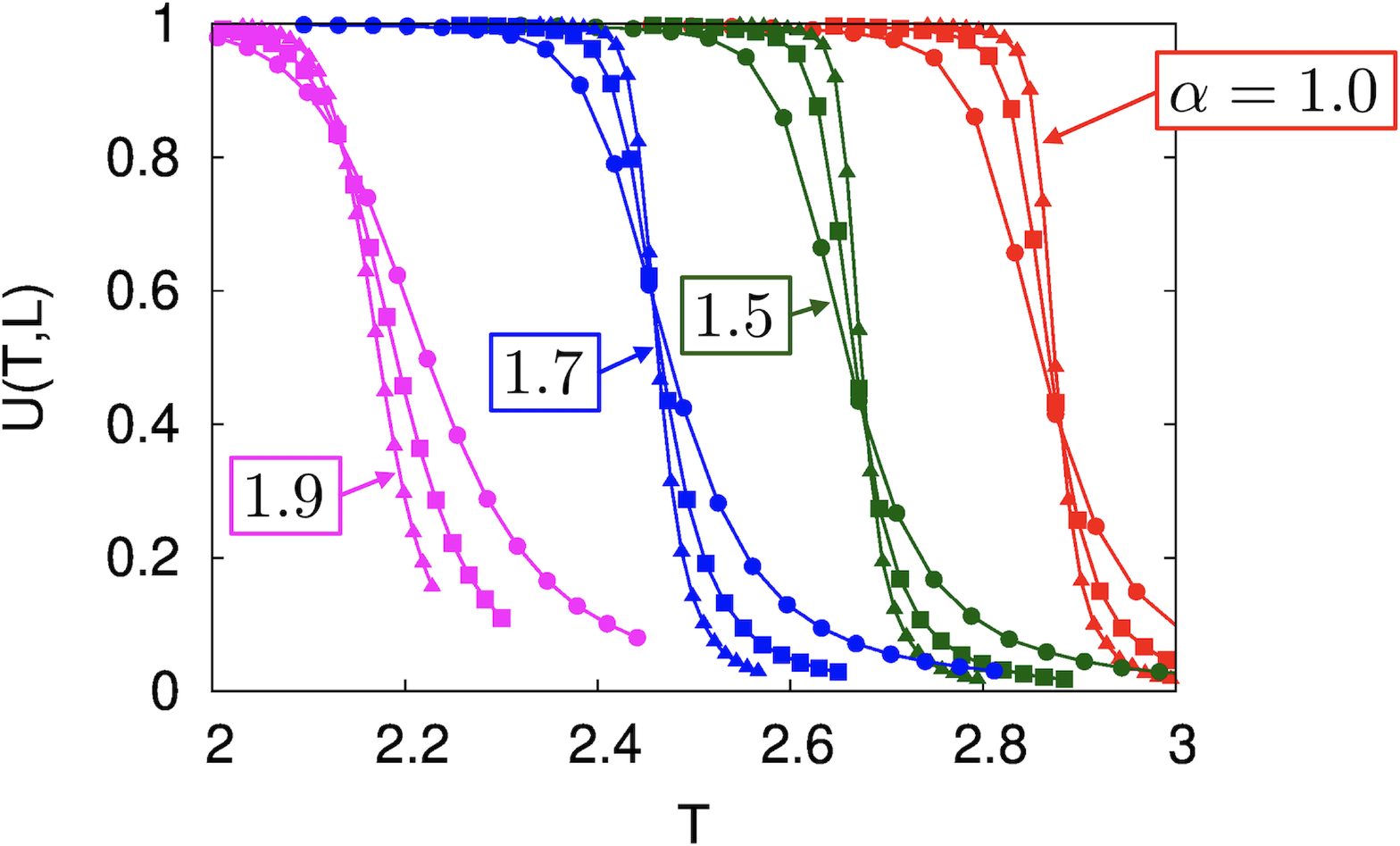}
\end{center}
\caption{(Color online) The temperature dependence of the Bider ratio defined by Eq.~(\ref{eqn:def_Binder}) 
for $\alpha=1.0$ (red), $1.5$ (green), $1.7$ (blue), and $1.9$ (magenta). 
The dimension $d$ is $1$ and the degree $k$ is $4$. The lattice size $L$ is 
$3,000$ (circles), $10,000$ (squares), and $30,000$ (triangles). 
The average is taken over $100$ different graphs.}
\label{fig:Binder1}
\end{figure}

\section{Result II: Application to a Ferromagnetic Ising Model}
\label{sec:result2}
To see how physical phenomena are affected by the geometric properties of the graph such as the exponent $\alpha$, 
we investigate a ferromagnetic Ising model on the graph. The reason why we choose 
this model as a test case is that the corresponding long-range 
model has been intensively studied until now. The Hamiltonian of the corresponding model is
\begin{equation}
{\cal H}=-\sum_{i < j} J_{ij}\sigma_i\sigma_j. 
\label{eqn:LRferro}
\end{equation}
The sum runs over all the pairs. The Ising spin $\sigma_i=\pm 1$ is located on a one-dimensional lattice of the size $L$ 
under the periodic boundary condition. The coupling constant $J_{ij}$ is given as 
\begin{equation}
J_{ij}=\frac{J}{d_{ij}^{\alpha}},
\end{equation}
where $J>0$ and $\alpha > 1$. The inequality $\alpha > 1$ is imposed to assure that the energy per spin converges to a finite value. 
The power-law decay in the coupling constant $J_{ij}$ corresponds to that in the probability that two vertices are connected by an edge 
(see Eq.~(\ref{eqn:Pgraph})). It is well established by previous 
studies\cite{AndersonYuval71, Cardy81, Aizenman88, ImbrieNewman88, LuijtenMessingfeld01, FukuiTodo09, 
Fisher72, Sak73, Kostalitz76, LuijtenBlote97} that this long-range model exhibits various critical 
behavior as the exponent $\alpha$ changes. When $1<\alpha \le 3/2$, the model belongs to a mean-field universality class. 
The critical exponents do not change in this range. When $3/2<\alpha<2$, the critical exponents continuously changes with $\alpha$. 
Finally, at $\alpha=2$, the model exhibits a marginal Kosterlitz-Thouless transition.

In this section, we study a ferromagnetic Ising model on our graph whose Hamiltonian is given as 
\begin{equation}
{\cal H}=-J\sum_{\langle ij \rangle} \sigma_i\sigma_j, 
\end{equation}
where $J>0$. We hereafter set $J/k_{\rm B}$ to $1$ for simplicity, where $k_{\rm B}$ is the Boltzmann constant. 
The sum runs over the nearest neighboring pairs of the graph. The dimension $d$ and the degree $k$ are set to $1$ 
and $4$, respectively. The maximum sizes we have investigated are $30,000$ for $\alpha \ne 2$ and $300,000$ for $\alpha = 2$. 
The graph is made by performing MCMC simulation of the method (c) (both the reverse and list-based update methods) 
for $10,000$~MCS. As we see from Fig.~\ref{fig:RelaxTime}, the autocorrelation time $\tau_{\rm graph}$ is much smaller 
than $10,000$~MCS for all the sizes $L$ and exponents $\alpha$ we have investigated. 
After each spin $\sigma_i$ is initially set to either $-1$ or $+1$ with the equal probability, 
the spin configuration is updated by the Swendsen-Wang cluster algorithm\cite{SwendsenWang87} for $20,000$~MCS. 
The first $10,000$~MCS is for equilibration and the subsequent $10,000$~MCS is for measurement. 
We have checked that $10,000$~MCS is enough for the system to be equilibrated in all the cases. 

\begin{figure}[t]
\begin{center}
\includegraphics[width=8cm]{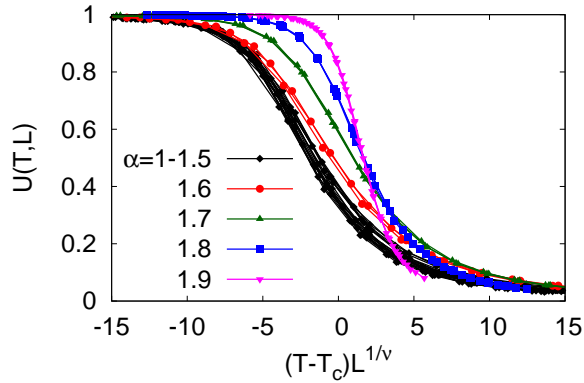}
\end{center}
\caption{(Color online) Scaling plot of the Binder ratio for $\alpha=1.0$, $1.1$, $1.2$, $\cdots$, $1.9$. 
The dimension $d$ is $1$ and the degree $k$ is $4$. The lattice size $L$ is $3,000$, $10,000$, and $30,000$. 
The data with $\alpha=1.0$, $1.1$, $\cdots$, $1.5$ are plotted by the same symbols (black diamonds). 
The average is taken over $100$ different graphs.}
\label{fig:Binder2}
\end{figure}

Figure~\ref{fig:Binder1} shows the temperature dependence of the Binder ratio\cite{Binder81} defined by 
\begin{equation}
U(T, L)=\frac{1}{2}\left(3-\frac{\langle M^4 \rangle}{\langle M^2 \rangle^2}\right), 
\label{eqn:def_Binder}
\end{equation}
where $M\equiv \sum_i \sigma_i$ is the magnetization and $\langle \cdots \rangle$ denotes the thermal average 
over the Boltzmann distribution. The exponent $\alpha$ is $1.0$, $1.5$, $1.7$, and $1.9$ from right to left. 
Because the Binder ratio is dimensionless, its finite-size scaling form is
\begin{equation}
U(T, L)=F_{\rm B}((T-T_c)L^{1/\nu}), 
\end{equation}
where $T_{\rm c}$ is a critical temperature, $\nu$ is a critical exponent which characterizes the divergence of the correlation lengh, 
and $F_{\rm B}(x)$ is a scaling function of the Binder ratio. Therefore, we can estimate the critical temperature $T_{\rm c}$ 
from the crossing point of $U(T,L)$'s for different sizes. We see from Fig.~\ref{fig:Binder1} that the critical temperature decreases 
with increasing $\alpha$. Figure~\ref{fig:Binder2} shows the result of scaling analysis of the Binder ratio for $\alpha=1.0$, 
$1.1$, $1.2$, $\cdots$, $1.9$. In the scaling analysis, we used a method based on Bayesian inference~\cite{Harada11, Harada15}. 
The data with $\alpha=1.0$, $1.1$, $\cdots$, $1.5$, which are expected to belong to a common mean-field universality class, 
are plotted by the same symbols (black diamonds). We see that these data roughly collapse into a single scaling curve. 
We also notice that scaling function continuously changes as the exponent $\alpha$ increases from $1.5$ to $1.9$. 
A sign of this continuous change is also found in Fig.~\ref{fig:Binder1}, i.e., the value of Binder ratio at the crossing point 
increases as $\alpha$ increases from $1.5$ to $1.9$. This behavior is consistent with the fact that the universality class 
of the corresponding long-range model changes continuously for $1.5 < \alpha < 2.0$. 
We have estimated $T_{\rm c}$ and $1/\nu$ through the scaling analysis. 

\begin{figure}[t]
\begin{center}
\includegraphics[width=8cm]{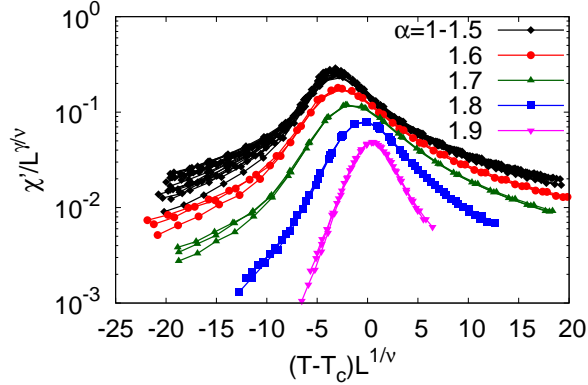}
\end{center}
\caption{(Color online) Scaling plot of the finite-lattice magnetic susceptibility $\chi'$ defined by Eq.~(\ref{eqn:def_Dchi}) 
for $\alpha=1.0$, $1.1$, $1.2$, $\cdots$, $1.9$. The dimension $d$ is $1$ and the degree $k$ is $4$. 
The lattice size $L$ is $3,000$, $10,000$, and $30,000$. The data with $\alpha=1.0$, $1.1$, $\cdots$, $1.5$ 
are plotted by the same symbols (black diamonds). The average is taken over $100$ different graphs.}
\label{fig:Chi}
\end{figure}

\begin{figure}[t]
\begin{center}
\includegraphics[width=8cm]{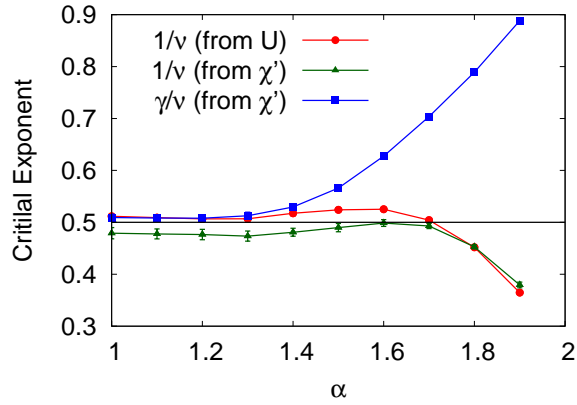}
\end{center}
\vspace{5mm}
\caption{(Color online) $1/\nu$ and $\gamma/\nu$ estimated through the scaling analysis are plotted as a function of $\alpha$. 
$1/\nu$ estimated from the Binder ratio $U$ and that from the susceptibility $\chi'$ are denoted by 
circles and triangles, respectively.}
\label{fig:Cexponent}
\end{figure}

We also measured  a finite-lattice magnetic susceptibility\cite{FS_susceptibility} defined by
\begin{equation}
\chi'(T,L) \equiv \frac{1}{NT}\left( \langle M^2 \rangle -\langle |M| \rangle^2 \right),
\label{eqn:def_Dchi}
\end{equation}
where $N$ is the number of spins. Since $d=1$, $N$ is equal to $L$. The scaling form of $\chi'$ is
\begin{equation}
\chi'(T,L)=L^{\gamma/\nu}F_{\chi'}((T-T_c)L^{1/\nu}), 
\end{equation}
where $\gamma$ is a critical exponent which characterizes the divergence of the magnetic susceptibility. 
The result of scaling analysis of $\chi'$ is shown in Fig.~\ref{fig:Chi}. We again see that scaling function for $\alpha>3/2$ 
continuously changes with $\alpha$. We have estimated $T_{\rm c}$, $1/\nu$, and $\gamma/\nu$ through the scaling analysis. 
In Fig.~\ref{fig:Cexponent}, $1/\nu$ and $\gamma/\nu$ estimated by the scaling analysis are plotted as a function of $\alpha$. 
Both $1/\nu$ and $\gamma/\nu$ are $1/2$ in the mean-field universality class. 
We see that  $1/\nu$ and $\gamma/\nu$ are close to $1/2$ for $\alpha \le 3/2$ and they continuously changes for $\alpha > 3/2$. 

\begin{figure}[t]
\begin{center}
\includegraphics[width=8cm]{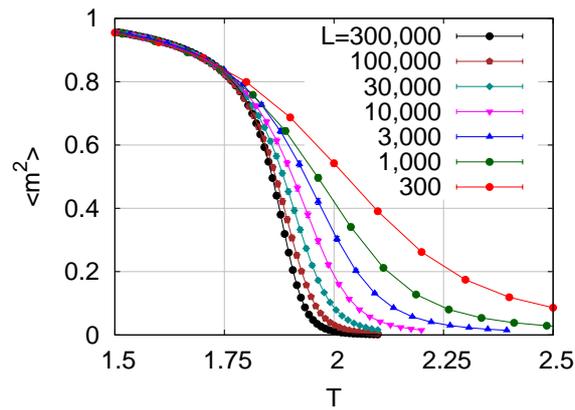}
\end{center}
\vspace{5mm}
\caption{(Color online) The temperature dependence of $\langle m^2 \rangle$, where $m\equiv \sum_i\sigma_i /N$ 
and $\langle \cdots \rangle$ denotes the thermal average. The dimension $d$ is $1$, the degree $k$ is $4$, 
and the exponent $\alpha$ is $2.0$. The average is taken over $100$ different graphs.}
\label{fig:M2_a20}
\end{figure}

\begin{figure}[t]
\begin{center}
\includegraphics[width=8cm]{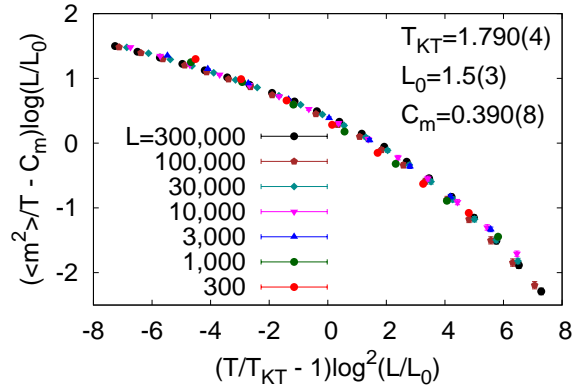}
\end{center}
\vspace{5mm}
\caption{(Color online) Scaling plot of $\langle m^2 \rangle$ for the data shown in Fig.~\ref{fig:M2_a20}. 
The assumed scaling form is given by Eq.~(\ref{eqn:KTscaling}). The three parameters $T_{\rm KT}$, 
$L_0$, and $C_{\rm m}$ in Eq.~(\ref{eqn:KTscaling}) are estimated by fitting the data. 
}
\label{fig:M2_a20_scaling}
\end{figure}

We next show the data for $\alpha=2.0$. As mentioned before, the corresponding long-range model with the Hamiltonian of 
Eq.~(\ref{eqn:LRferro}) exhibits a Kosterlitz-Thouless transition. In Fig.~\ref{fig:M2_a20}, we plot $\langle m^2 \rangle$ for $\alpha=2.0$ 
as a function of temperature, where $m\equiv \sum_i\sigma_i /N$. 
We see that the data exhibit a strong size dependence, which is one of characteristics of the Kosterlitz-Thouless transition. 
We also find that $\langle m^2 \rangle$ drops to zero quite rapidly as the size increases. This behavior indicates 
the existence of a universal jump~\cite{NelsonKosterlitz77, Aizenman88, HaradaKawashima97, HaradaKawashima98, LuijtenMessingfeld01, FukuiTodo09} 
which is ubiquitous in the Kosterlitz-Thouless transition. 

We also perform a finite-size scaling analysis of the data. In the scaling analysis, we assume the following 
scaling form~\cite{KosterlitzThouless74, WeberMinnhagen87, HaradaKawashima97, HaradaKawashima98, FukuiTodo09} suggested by renormalization group equations 
of the Kosterlitz-Thouless transition: 
\begin{equation}
\frac{\langle m^2 \rangle}{T}=C_{\rm m}+\ell^{-1} F(t\ell^2), 
\label{eqn:KTscaling}
\end{equation}
where $F(x)$ is a scaling function, $\ell \equiv L/L_0$, $t\equiv T/T_{\rm KT}$, $T_{\rm KT}$ is a Kosterlitz-Thouless 
transition temperature, and $L_0$ and $C_{\rm m}$ are constants. Although the constant $C_{\rm m}$ 
in Eq.~(\ref{eqn:KTscaling}) has been set to $0.5$ in the scaling analysis of the corresponding long-range model 
with the Hamiltonian of Eq.~(\ref{eqn:LRferro})~\cite{FukuiTodo09}, we treat $C_{\rm m}$ as a fitting parameter. The result of 
scaling analysis is shown in Fig.~\ref{fig:M2_a20_scaling}. We see that the scaling works nicely. 
From the scaling analysis, we estimate $T_{\rm KT}$ to be $1.790(4)$.

\begin{figure}[t]
\begin{center}
\includegraphics[width=8cm]{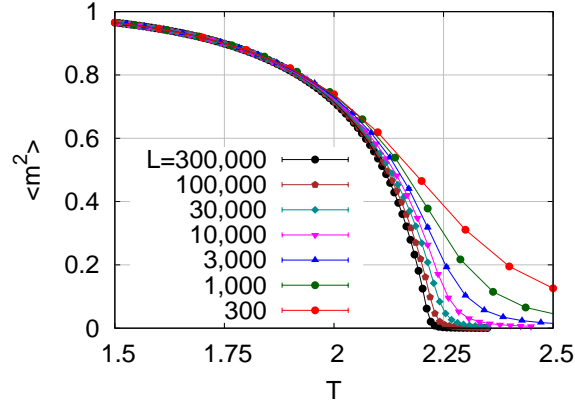}
\end{center}
\vspace{5mm}
\caption{(Color online) The temperature dependence of $\langle m^2 \rangle$ for initial graphs. 
The dimension $d$ is $1$, the degree $k$ is $4$, and the exponent $\alpha$ is $2.0$. 
The average is taken over $100$ different graphs.}
\label{fig:M2_a20_init_graph}
\end{figure}

Lastly, Fig.~\ref{fig:M2_a20_init_graph} shows the temperature dependence of $\langle m^2 \rangle$ for initial graphs. 
Initial graphs are prepared only by the method described in Sect.~\ref{subsec:initial_graph}, and we do not perform 
a subsequent MCMC simulation to update the graph. The exponent $\alpha$ is $2.0$. 
We have shown the probability distribution $P(r)$ for initial graphs in Fig.~\ref{fig:Hist} by blue line. 
On the other hand, $P(r)$ calculated from graphs used in Fig.~\ref{fig:M2_a20} has been shown in Fig.~\ref{fig:Hist} 
by red line. We see that the temperature dependence of $\langle m^2 \rangle$ 
in Fig.~\ref{fig:M2_a20_init_graph} is significantly different from that in Fig.~\ref{fig:M2_a20}. 
It should be noted that the ranges of temperature are the same in the two figures. 
This result tells us that, to obtain correct data, we should {\it not} use initial graphs, 
although they approximately satisfy the desired distribution given by Eq.~(\ref{eqn:Pgraph}). 

\section{Conclusions}
\label{sec:conclusions}
In this paper, we consider a geometric graph which satisfies the following two conditions: 
(i) The degree of each vertex is fixed to a positive integer $k$. 
(ii) The probability that two vertices located on a $d$-dimensional hypercubic lattice are connected 
by an edge is proportional to $d_{ij}^{-\alpha}$, where $d_{ij}$ is the distance between the two vertices 
and $\alpha$ is a positive exponent. By changing $\alpha$, we can continuously change the graph from a lattice 
in a $d$-dimensional space\cite{note1}  ($\alpha=\infty$) to a random graph with a fixed connectivity ($\alpha=0$). 
Therefore, we can systematically examine how physical phenomena is affected by the graph structure of the system. 
Furthermore, because the degree distribution is fixed (see the condition (i)), we can deny the possibility that 
the change in the physical phenomena is caused by a change in the degree distribution. This is an important 
point for some problems like constraint satisfaction problems which are considered to sensitively depend on the degree distribution. 
To make this geometric graph, we have introduced a MCMC method which consists of two update methods, 
i.e., reverse update method and list-based update method. Because these two update methods 
work complementarily, we can efficiently update the graph by the MCMC method. 
We have also investigated a ferromagnetic Ising model defined on the geometric graph as a test case. 
As a result, we have confirmed that the nature of ferromagnetic transition significantly depends on the exponent $\alpha$. 
We hope that this work will stimulate further researches to reveal the relation between physical phenomena 
and the graph structure of the system. 

\begin{acknowledgment}
The author would like to thank Dr. Harada for fruitful discussion, especially for useful suggestion 
on the finite-size scaling analysis of the Kosterlitz-Thouless transition. 
\end{acknowledgment}

\appendix 
\section{Derivation of Eq.~(\ref{eqn:Pgen_reverse})}
\label{sec:appendix1}
\begin{figure}[t]
\begin{center}
\includegraphics[width=8cm]{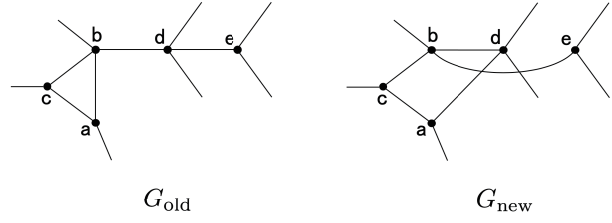}
\end{center}
\caption{Subgraphs of two graphs $G_{\rm old}$ and $G_{\rm new}$. The two graphs are the same except the subgraphs.}
\label{fig:loop1}
\end{figure}
We consider two graphs $G_{\rm old}$ and $G_{\rm new}$. Figure~\ref{fig:loop1} shows their subgraphs which 
contain five vertices. We assume that the two graphs are the same except the subgraphs. In order that $G_{\rm new}$ is 
generated from $G_{\rm old}$ by the five steps shown in Sect.~\ref{subsec:RUM}, a path of the length $l$ chosen by the steps 
should satisfy the following four conditions~\cite{note2}: 
\begin{itemize}
\item[(i)] The starting vertex $i_0$ is $a$. 
\item[(ii)] $i_1$ is $b$. 
\item[(iii)] $i_{l-1}$ is $d$. 
\item[(iv)] The end vertex $i_{l}$ is $e$. 
\end{itemize}

We hereafter consider the case that we do not perform the step (d), and show that 
it makes Eq.~(\ref{eqn:Pgen_reverse}) invalid. If the maximum path length $l_{\rm max}$ in step (b) is large enough, 
there are several paths which satisfy the above-mentioned four conditions
such as $p(a, b, d, e)$, $p(a, b, c, a, b, d, e)$, and so forth, where $p(\cdots)$ denotes a path and the vertices visited 
in the path are denoted in order in the parentheses. It should be noted that the second path $p(a, b, c, a, b, d, e)$ 
is excluded if we perform the step (d). The probability 
$P_{\rm gen}^{\rm R}(G_{\rm old}\rightarrow G_{\rm new})$ that a graph $G_{\rm new}$ is generated from $G_{\rm old}$ 
is written as
\begin{equation}
P_{\rm gen}^{\rm R}(G_{\rm old}\rightarrow G_{\rm new})=\sum_{p\in {\cal S}_p(G_{\rm old}\rightarrow G_{\rm new})}Q(p; G_{\rm old}),
\end{equation}
where $ {\cal S}_p(G_{\rm old}\rightarrow G_{\rm new})$ is the set of paths which satisfy the four conditions and 
$Q(p; G)$ is the probability that a path $p$ is chosen on the graph $G$. 
Because we consider the case that the degree of every vertex is $k$, $Q(p; G_{\rm old})$ is written as
\begin{equation}
Q(p; G)=\frac{1}{N} \frac{1}{k}\left(\frac{1}{k-1}\right)^{l(p)-1},
\label{eqn:Qestimate}
\end{equation}
where $l(p)$ is the length of the path $p$. The first factor in the right hand of Eq.~(\ref{eqn:Qestimate}) is the probability that $i_0$ is chosen 
among the $N$ vertices, the second factor is the probability that $i_1$ is chosen among the $k$ neighboring vertices of $i_0$, 
and the third factor is the probability that $i_s$ ($2\le s \le l$) is chosen among the $k$ neighboring vertices of $i_{s-1}$ 
except $i_{s-2}$. We see from Eq.~(\ref{eqn:Qestimate}) that $Q(p; G)$ only depends on $l(p)$. 
The probability $P_{\rm gen}^{\rm R}(G_{\rm new}\rightarrow G_{\rm old})$ is written in a similar way. 
The set ${\cal S}_p(G_{\rm new}\rightarrow G_{\rm old})$ contains paths such as 
$p(a, d, b, e)$ and $p(a, d, b, c, a, d, b, e)$. Now the point is that $Q(p(a, b, c, a, b, d, e); G_{\rm old})$  and 
$Q(p(a, d, b, c, a, d, b, e); G_{\rm new})$ are different because the lengths of the two paths are different. 
This difference is caused by the fact that the lengths of the two loops $p(a, b, c, a)$ and $p(a, d, b, c, a)$ 
contained in the two paths are different. As a result, 
Eq.~(\ref{eqn:Pgen_reverse}) becomes invalid. However, if we perform the step (d), Eq.~(\ref{eqn:Pgen_reverse}) becomes valid 
because these two paths are excluded in the step (d). 

\begin{figure}[t]
\begin{center}
\includegraphics[width=8cm]{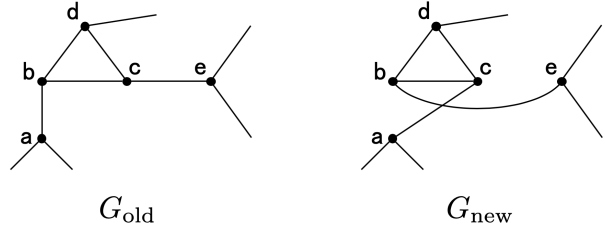}
\end{center}
\caption{An example of the case that a path has a loop which contains neither the starting nor end point. 
A path $p(a, b, c, d, b, c, e)$ in the graph $G_{\rm old}$ has a loop $p(b, c, d, b)$ which contains neither 
the starting point $a$ nor the end point $e$. The corresponding path in the reverse process is 
$p(a, c, b, d, c, b, e)$ in $G_{\rm new}$. The probabilities for the two paths to be chosen are the same 
because the lengths of the two paths are the same (see Eq.~(\ref{eqn:Qestimate})).}
\label{fig:loop2}
\end{figure}

The example shown in the previous paragraph tells us that, if a path chosen in the graph $G_{\rm old}$ has 
a loop which contains either the starting or end point, there is no corresponding path in the reverse process 
generating from $G_{\rm new}$ to $G_{\rm old}$, as the path $p(a, b, c, a, b, d, e)$ does not correspond to the path 
$p(a, d, b, c, a, d, b, e)$ in the reverse process. We therefore exclude such a path in the step (d). On the other hand, 
if the path has a loop which contains neither the starting nor end point, there is a corresponding path in the reverse process. 
As an example, we consider two graphs $G_{\rm old}$ and $G_{\rm new}$ which have the subgraphs shown 
in Fig.~\ref{fig:loop2}. 
We again assume that the two graphs are the same except the subgraphs. In this case, a path $p(a, b, c, d, b, c, e)$ 
in the graph $G_{\rm old}$ corresponds to a path $p(a, c, b, d, c, b, e)$ in $G_{\rm new}$. 
It is worth noticing that the two paths have a common loop. Because the lengths of the two paths 
are the same, the probabilities for the paths to be chosen are also the same. Therefore, Eq.~(\ref{eqn:Pgen_reverse}) 
is valid even if we allow these paths to be chosen. This is the reason why we do not exclude paths which have a loop 
containing neither the starting nor end point.

\section{Estimation of the Acceptance Probability in the List-Based Update Method}
\label{sec:appendix2}
As we mentioned in Sect.~\ref{subsec:LB}, Eq.~(\ref{eqn:LB11}) becomes invalid for $\alpha \ge d$ 
because Eq.~(\ref{eqn:LB7}) does. Therefore, we first explain the reason why Eq.~(\ref{eqn:LB7}) 
becomes invalid. If we take an average of $N_{\rm v}(G; \bm{r})$ over graphs, Eq.~(\ref{eqn:LB7}) 
is always valid. However, when $\alpha \ge d$ and $N_{\rm v}(G; \bm{r})$ 
rapidly decreases with increasing $|\bm{r}|$, the mean of $N_{\rm v}(G; \bm{r})$ 
can be much smaller than $1$ for large $\bm{r}$. Equation~(\ref{eqn:LB6}) becomes invalid in such a case 
because $N_{\rm v}(G; \bm{r})$, which is the number of edges of a graph $G$ 
which belong to the vector $\bm{r}$, is a non-negative integer. 

To estimate the condition for Eq.~(\ref{eqn:LB7}) being invalid, we calculate the mean of  
$N_{\rm v}(G; \bm{r})$. To this end, we introduce the probability $p(r)$ that two vertices separated 
by a distance $r$ is connected by an edge. The mean of $N_{\rm v}(G; \bm{r})$ and $p(r)$ are 
related as
\begin{equation}
\overline{N_{\rm v}(G; \bm{r})} = N p(|\bm{r}|),
\label{eqn:Nv_and_p}
\end{equation}
where $\overline{\cdots}$ denotes the average over graphs. On the other hand, because the degree of 
each vertex is $k$, we obtain
\begin{equation}
\sum_{\bm{r}} p(|\bm{r}|) \sim \int_{1}^{L} {\rm d}r r^{d-1} p(r)=k, 
\label{eqn:pr_condition}
\end{equation}
where we have used the lattice constant as a unit of $r$. By substituting 
\begin{equation}
p(r)=c^*r^{-\alpha},
\label{eqn:powerlaw_pr}
\end{equation}
into (\ref{eqn:pr_condition}), the proportionality constant $c^*$ is roughly estimated as 
\begin{equation}
c^*\sim\left\{
\begin{array}{cc}
L^{\alpha-d} & (\alpha < d), \\
1/\log(L) & (\alpha = d), \\
1 & (\alpha >d).
\end{array}
\right.
\label{eqn:estimation_c*}
\end{equation}
From the relation $N=L^{d}$ and Eqs.~(\ref{eqn:Nv_and_p}), (\ref{eqn:powerlaw_pr}), and (\ref{eqn:estimation_c*}), 
we find that $\overline{N_{\rm v}(G; \bm{r})}$ for $|\bm{r}|\sim L$ is much smaller 
than $1$ if $\alpha \ge d$.

\begin{figure}[t]
\begin{center}
\includegraphics[width=8cm]{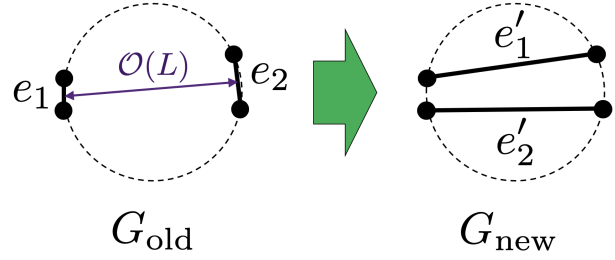}
\end{center}
\caption{(Color online) Schematic illustration of typical events in the list-based update method 
when $\alpha > d$. One-dimensional lattice under the periodic boundary condition is 
depicted by a dashed circle. When $\alpha > d$, the lengths of the two edges $e_1$ and $e_2$ 
are much smaller than $L$ in most cases. On the other hand, the lengths of the two new edges 
$e_1'$ and $e_1'$ are typically of the order of $L$ because the two edges $e_1$ and $e_2$ 
are chosen independently.}
\label{fig:smallAP}
\end{figure}

We next consider what typically happens when $\alpha > d$ and $L$ is large. 
It is schematically shown in Fig.~\ref{fig:smallAP}. 
We see from Eqs.~(\ref{eqn:Nv_and_p}), (\ref{eqn:powerlaw_pr}), and (\ref{eqn:estimation_c*}) that
$\overline{N_{\rm v}(G; \bm{r})}$ is smaller than $1$ if $|\bm{r}|$ is larger than a threshold length
\begin{equation}
L_{\rm th}=L^{d/\alpha}.
\label{eqn:def_Lth}
\end{equation} 
Because $\alpha > d > 0$ and $L$ is large, $L_{\rm th}$ is much smaller than $L$. 
Therefore, we can assume that the lengths of the two edges $e_1$ and $e_2$ chosen in the list-based method 
are much smaller than $L$ in most cases. On the other hand, the two edges are separated 
by a distance of ${\cal O}(L)$ in most cases because the two edges are chosen independently. 
This means that the lengths of the two new edges $e_1'$ and $e_1'$ made in step (f) are also 
of the order of $L$. Now the point is that $N_{\rm v}$'s in Eq.~(\ref{eqn:LB4}) are at least one 
because there are two edges $e_1'$ and $e_2'$ in the graph $G_{\rm new}$. 
Therefore, the actual value of $T_1(G_{\rm new}\rightarrow G_{\rm int})$ 
is much smaller than the approximate estimation of Eq.~(\ref{eqn:LB9}) obtained 
by replacing $N_{\rm v}$'s in Eq.~(\ref{eqn:LB4}) with their mean values given by Eq.~(\ref{eqn:LB7}). 
As a result, the acceptance probability given by Eq.~(\ref{eqn:LB5}) becomes much smaller than 
its approximate estimation of Eq.~(\ref{eqn:LB11}). 

Lastly, we roughly estimate the mean of the acceptance probability. When two edges $e_1$ and $e_2$ 
are separated by a distance which is much larger than their lengths, the two edges are approximately 
regarded as two points on the scale of the distance between the two edges (see Fig.~\ref{fig:smallAP}). 
Therefore, we hereafter regard the two edges as two points and denote their distance by $r_{\rm e}$. 
In this approximation, the lengths of the two new edges $e_1'$ and $e_2'$ are $r_{\rm e}$. 
We now reconsider the approximate estimations of Eqs.~(\ref{eqn:LB8}) and (\ref{eqn:LB9}). 
We can regard the first approximation Eq.~(\ref{eqn:LB8}) as valid because the lengths of the two edges 
$e_1$ and $e_2$ are less than $L_{\rm th}$ given by Eq.~(\ref{eqn:def_Lth}) in most cases. 
However, we have to modify the second approximate estimation of Eq.~(\ref{eqn:LB9}) 
because $r_{\rm e}$ often becomes larger than $L_{\rm th}$. 
By replacing $N_{\rm v}$'s in Eq.~(\ref{eqn:LB4}) with $C^*r_{\rm e}^{-\alpha}$ for $r_{\rm e} < L_{\rm th}$, 
or $1$ for $r_{\rm e} > L_{\rm th}$, we obtain
\begin{equation}
T_1(G_{\rm new}\rightarrow G_{\rm int}) \approx \left\{
\begin{array}{cc}
\frac{2r_{\rm e}^{2\alpha}}{(N_{\rm nz}^*C^*)^2} & (r_{\rm e}<L_{\rm th}), \\
\frac{2}{(N_{\rm nz}^*)^2} & (r_{\rm e} > L_{\rm th}). 
\end{array}
\right.
\label{eqn:T1estimation2}
\end{equation}
By substituting Eqs.~(\ref{eqn:LB8}), (\ref{eqn:LB10}) with $|\bm{r}_1'|=|\bm{r}_2'|=r_{\rm e}$, 
and (\ref{eqn:T1estimation2}) into Eq.~(\ref{eqn:LB5}), we obtain
\begin{equation}
A^{\rm LB}(G_{\rm new}\rightarrow G_{\rm new}) \approx \left\{
\begin{array}{cc}
1 & (r_{\rm e}<L_{\rm th}), \\
(C^*r_{\rm e}^{-\alpha})^2 & (r_{\rm e} > L_{\rm th}).
\end{array}
\right.
\label{eqn:ALBestimation2}
\end{equation}
From Eq.~(\ref{eqn:Nv_and_p}), we see that $C^*$ in Eq.~(\ref{eqn:LB7}) is related to 
$c^*$ in Eq.~(\ref{eqn:powerlaw_pr}) as $C^*=Nc^*$. Therefore, 
$C^*$ is $N$ when $\alpha > d$ (see Eq.~(\ref{eqn:estimation_c*})). We also 
find from Eq.~(\ref{eqn:def_Lth}) that $N=(L_{\rm th})^{\alpha}$. 
By substituting them into (\ref{eqn:ALBestimation2}), we obtain
\begin{equation}
A^{\rm LB}(G_{\rm new}\rightarrow G_{\rm new}) \approx \left\{
\begin{array}{cc}
1 & (r_{\rm e}<L_{\rm th}), \\
\left(\frac{r_{\rm e}}{L_{\rm th}}\right)^{-2\alpha} & (r_{\rm e} > L_{\rm th}). 
\end{array}
\right.
\label{eqn:ALBestimation3}
\end{equation}
This equation means that the acceptance probability depends on the distance $r_{\rm e}$ 
between the two edges $e_1$ and $e_2$. Therefore, to calculate the mean of the acceptance probability, 
we have to take an average over $r_{\rm e}$. Let us denote the probability density 
of $r_{\rm e}$ by $P_{\rm e}(r_{\rm e})$. Because the two edges are chosen independently, 
$P_{\rm e}(r_{\rm e})$ is approximately evaluated as 
\begin{equation}
P_{\rm e}(r_{\rm e})\approx \frac{1}{N}(r_{\rm e})^{d-1}. 
\end{equation}
The mean of the acceptance probability is written with $P_{\rm e}(r)$ as 
\begin{equation}
\overline{A^{\rm LB}(G_{\rm new}\rightarrow G_{\rm new})}
\approx\int_{1}^{L} {\rm d}r_{\rm e} P_{\rm e}(r_{\rm e})A^{\rm LB}(G_{\rm new}\rightarrow G_{\rm new}).
\label{eqn:def_ave_ALB}
\end{equation}
By substituting Eq.~(\ref{eqn:ALBestimation3}) into (\ref{eqn:def_ave_ALB}) and calculating the integral, 
we finally obtain
\begin{equation}
\overline{A^{\rm LB}(G_{\rm new}\rightarrow G_{\rm new})}
\sim \frac{(L_{\rm th})^{d}}{N}=N^{d/\alpha-1},
\end{equation}
where we have used the fact that $d-2\alpha < d-\alpha < 0$ to evaluate the integral, 
and Eq.~(\ref{eqn:def_Lth}) to go from the second line to the third. 
As mentioned before, the validity of this rough estimation has been 
confirmed numerically in Fig.~\ref{fig:Paccepted}~(ii).


\begin{thebibliography}{99} 

\bibitem{SherringtonKirkpatrick75}
D. Sherrington and S. Kirkpatrick, Phys. Rev. Lett. {\bf 35}, 1792 (1975). 

\bibitem{SherringtonKirkpatrick78}
D. Sherrington and S. Kirkpatrick, Phys. Rev. B {\bf 17}, 4384 (1978). 

\bibitem{Parisi80}
G. Parisi, J. Phys. A {\bf 13}, L115, 1101, and 1887 (1980). 

\bibitem{AndersonYuval71}
P. W. Anderson and G. Yuval, J. Phys. C {\bf 4}, 607 (1971).

\bibitem{Cardy81}
J. L. Cardy, J. Phys. A {\bf 14}, 1407 (1981). 

\bibitem{Aizenman88}
M. Aizenman, J. T. Chayes, L. Chayes, and C. M. Newman, J. Stat. Phys. {\bf 50}, 1 (1988).

\bibitem{ImbrieNewman88}
J. Z. Imbrie and C. M. Newman, Commun. Math. Phys. {\bf 118}, 303 (1988).

\bibitem{LuijtenMessingfeld01}
E.~Luijten and H.~Me{\ss}ingfeld, Phys. Rev. Lett. {\bf 86}, 5305 (2001).

\bibitem{FukuiTodo09}
K. Fukui and S. Todo, J. Comput. Phys. {\bf 228}, 2629 (2009). 

\bibitem{BrayMooreYoung86}
A. J. Bray, M. A. Moore, and A. P. Young, Phys. Rev. Lett. {\bf 56}, 2641 (1986).

\bibitem{FisherHuse88}
D. S. Fisher and D. A. Huse, Phys. Rev. B {\bf 38}, 386 (1988).

\bibitem{KatzgraberYoung03}
H. G. Katzgraber and A. P. Young, Phys. Rev. B {\bf 67}, 134410 (2003). 

\bibitem{KatzgraberYoung05}
H. G. Katzgraber and A. P. Young, Phys. Rev. B {\bf 72}, 184416 (2005). 

\bibitem{Kleinberg00}
J. M. Kleinberg, Nature {\bf 406}, 845 (2000).

\bibitem{Coppersmith02}
D. Coppersmith, D. Gamarnik, and M. Sviridenko, Random Struct. Algor. {\bf 21}, 1 (2002).

\bibitem{Biskup04}
M. Biskup, Ann. Probab. {\bf 32}, 2938 (2004). 

\bibitem{Leuzzi08}
L. Leuzzi, G. Parisi, F. Ricci-Tersenghi, and J. J. Ruiz-Lorenzo, Phys. Rev. Lett. {\bf 101}, 107203 (2008). 

\bibitem{Katzgraber09}
H. G. Katzgraber, D. Larson, and A. P. Young, Phys. Rev. Lett. {\bf 102}, 177205 (2009). 

\bibitem{Walker77}
A.~J.~Walker, ACM Trans. Math. Software {\bf 3}, 253 (1977).

\bibitem{Lin65}
S. Lin, Bell Syst. Tech. J. {\bf 44}, 2245 (1965). 

\bibitem{autocorrelation_time}
D. P. Landau and K. Binder, {\it A Guide to Monte Carlo Simulations in Statistical Physics} 
(Cambridge University Press, Cambridge, U.K., 2015) 4th ed., p.~93. 

\bibitem{Fisher72}
M.~E.~Fisher, S.-k.~Ma, and B.~G.~Nickel, Phys. Rev. Lett. {\bf 29}, 917 (1972). 

\bibitem{Sak73}
J.~Sak, Phys. Rev. B {\bf 8}, 281 (1973).

\bibitem{Kostalitz76}
J.~M.~Kosterlitz, Phys. Rev. Lett. {\bf 37}, 1577 (1976).

\bibitem{LuijtenBlote97}
E.~Luijten and H.~W.~J.~Bl{\"o}te, Phys. Rev. B {\bf 56}, 8945 (1997).

\bibitem{SwendsenWang87}
R.~H.~Swendsen and J.-S.~Wang, Phys. Rev. Lett. {\bf 58}, 86 (1987). 

\bibitem{Binder81} 
K. Binder, Z. Phys. B {\bf 43}, 119 (1981).

\bibitem{Harada11}
K. Harada, Phys. Rev. E {\bf 84}, 056704 (2011).

\bibitem{Harada15}
K. Harada, Phys. Rev. E {\bf 92}, 012106 (2015). 

\bibitem{FS_susceptibility}
The book in Ref.~\citen{autocorrelation_time}, p.~84. 

\bibitem{NelsonKosterlitz77}
D. R. Nelson and  J. M. Kosterlitz, Phys. Rev. Lett. {\bf 39}, 1201 (1977). 

\bibitem{HaradaKawashima97}
K. Harada and N. Kawashima, Phys. Rev. B {\bf 55}, R11949 (1997).

\bibitem{HaradaKawashima98}
K. Harada and N. Kawashima, J. Phys. Soc. Jpn. {\bf 67}, 2768 (1998). 

\bibitem{KosterlitzThouless74}
J. M. Kosterlitz and D. J. Thouless, J. Phys. C {\bf 7}, 1046 (1974).

\bibitem{WeberMinnhagen87}
H. Weber and P. Minnhagen, Phys. Rev. B {\bf 37}, 5986 (1987). 

\bibitem{note1}
The graph becomes a $d$-dimensional hypercubic lattice when $k=2d$ and $\alpha=\infty$. 

\bibitem{note2}
$G_{\rm new}$ is also generated from a reverse path which satisfies the following four conditions: 
(i) $i_0$ is $e$. (ii) $i_1$ is $d$. (iii) $i_{l-1}$ is $b$. (iv) $i_l$ is $a$. However, this fact does not affect 
the argument below. 

\end{thebibliography}
\end{document}